\documentclass[journal,10pt,twocolumn]{IEEEtran}
\makeatletter
 \def\@textbottom{\vskip \z@ \@plus 1pt}
 \let\@texttop\relax
 
\makeatother
\usepackage{fancyhdr}
\usepackage{graphicx}
\usepackage[utf8]{inputenc}
\usepackage{amsmath}
\usepackage{amsfonts}
\usepackage{amssymb}
\usepackage{algorithm}
\usepackage{algorithmic}
\usepackage{amsthm}

\usepackage{cite}
\usepackage{epsf,bm}
\usepackage{epstopdf}
\usepackage{xcolor}
\usepackage{color}
\usepackage{hyperref}
\usepackage{breqn}
\newtheorem{theorem}{Theorem}
\makeatletter
\def\BSTATE{\STATE\hskip-\ALG@thistlm}
\makeatother

\begin{document}

\title{Optimization of Energy-Constrained Wireless Powered Communication Networks with Heterogeneous Nodes}
\author{\large Mohamed A. Abd-Elmagid$^*$, Tamer ElBatt$^*$$^\dagger$, and Karim G. Seddik$^\ddagger$ \\ [.05in]
\small  \begin{tabular}{c} $^*$Wireless Intelligent
Networks Center (WINC), Nile University, Giza, Egypt.\\$^\dagger$Dept. of EECE, Faculty of Engineering, Cairo University, Giza, Egypt. \\
$^\ddagger$Electronics and Communications Engineering Department, American University, AUC Avenue, New Cairo 11835, Egypt.\\

email: m.abdelaziz@nu.edu.eg, telbatt@ieee.org, kseddik@aucegypt.edu
\end{tabular} }
%\author{\large Mohamed A. Abd-Elmagid, Tamer ElBatt, and Karim G. Seddik}

\maketitle
\begin{abstract}
In this paper, we generalize conventional time division multiple access (TDMA) wireless networks to a new type of wireless networks coined generalized wireless powered communication networks (g-WPCNs). Our prime objective is to optimize the design of g-WPCNs where nodes are equipped with radio frequency (RF) energy harvesting circuitries along with constant energy supplies. This constitutes an important step towards a generalized optimization framework for more realistic systems, beyond prior studies where nodes are solely powered by the inherently limited RF energy harvesting. Towards this objective, we formulate two optimization problems with different objective functions, namely, maximizing the sum throughput and maximizing the minimum throughput (maxmin) to address fairness. First, we study the sum throughput maximization problem, investigate its complexity and solve it efficiently using an algorithm based on alternating optimization approach. Afterwards, we shift our attention to the maxmin optimization problem to improve the fairness limitations associated with the sum throughput maximization problem. The proposed problem is generalized, compared to prior work, as it seemlessly lends itself to prior formulations in the literature as special cases representing extreme scenarios, namely, conventional TDMA wireless networks (no RF energy harvesting) and standard WPCNs, with only RF energy harvesting nodes. In addition, the generalized formulation encompasses a scenario of practical interest we introduce, namely, WPCNs with two types of nodes (with and without RF energy harvesting capability) where legacy nodes without RF energy harvesting can be utilized to enhance the system sum throughput, even beyond WPCNs with all RF energy harvesting nodes studied earlier in the literature. We establish the convexity of all formulated problems which opens room for efficient solution using standard techniques. Our numerical results show that conventional TDMA wireless networks and WPCNs with only RF energy harvesting nodes are considered as lower bounds on the performance of the generalized problem setting in terms of the maximum sum throughput and maxmin throughput. Moreover, the results reveal valuable insights and throughput-fairness trade-offs unique to our new problem setting.


\end{abstract}

\IEEEpeerreviewmaketitle
%
%
%*****************************************************************************************
\section{Introduction}
One of the significant challenges is to prolong the lifetime of energy-constrained wireless networks which are powered by finite capacity batteries. Although, the lifetime of such networks can be extended by replacing or recharging the batteries, it may be inconvenient and costly. Therefore, energy harvesting has been considered a promising technique to prolong the network's lifetime \cite{1,2} since it provides wireless devices with the capability of perpetual charging of their batteries through harvesting energy from the surrounding environment. In this context, mobile devices can harvest energy from different natural sources, e.g., solar, thermal, vibrational, electromagnetic, etc. \cite{3,4,5,6}.

 RF energy harvesting has recently become a growing research thrust enabled by the design of novel harvesting circuitries which allow wireless devices to continuously harvest energy from the ambient radio environment. Significant research has been conducted on interference alignment networks with wireless energy transfer \cite{inter1,inter2,inter3,inter4}. Exploiting the fact that RF signals bear, both, energy and information at the same time, a dynamic simultaneous wireless information and power transfer scheme called SWIPT has been proposed in \cite{7,8,9,10,11,12}. SWIPT was first studied from an information-theoretic perspective in \cite{7,8}. The fundamental trade-off between simultaneously transmitting information and harvesting energy is characterized for narrowband noisy channels in \cite{7} and for frequency-selective channels in \cite{8}. Afterwards, from a communication-theoretic perspective, the fundamental trade-off between transmitting energy and transmitting information over a point-to-point noisy link is studied in \cite{9}. Motivated by the fact that energy harvesting circuits are unable to harvest energy and decode information at the same time, the authors in \cite{10} proposed two practical receiver designs, namely, time switching and power splitting. For the time switching scheme, a receiving antenna periodically switches between the energy harvesting receiver and the information decoding receiver. On the other hand, for the power splitting scheme, the received signal is split into two streams with different power levels; one is sent to the energy harvesting receiver and the other to the information decoding receiver. In addition, \cite{11} introduced dynamic power splitting as a general SWIPT operation and proposed two practical SWIPT receiver architectures: 1) separated information and energy receivers and 2) integrated information and energy receivers. Moreover, SWIPT has been proposed and studied for orthogonal frequency division multiplexing (OFDM) systems in \cite{12}.

 Another line of research has recently considered RF-powered cognitive radio networks \cite{13,14,15} whereby the secondary users are assumed to have RF energy harvesting capability so that they could harvest energy whether from the RF primary users' signals or from other ambient RF sources. The amount of harvested energy is then used for data transmission. First, \cite{13} studied the optimal mode selection policy of whether the secondary users should harvest RF energy or access the spectrum in each slot time in order to maximize the expected total throughput. The optimal spectrum sensing policy was investigated in \cite{14} to maximize the expected total throughput subject to two constraints, namely, an energy causality constraint and a collision constraint. The former guarantees that the total consumed energy is less than or equal to the total harvested energy, while the collision constraint protects the primary user by guaranteeing a minimum QoS requirement. In \cite{15}, the optimal transmission power and density, for the cognitive nodes, were derived in order to maximize the secondary network throughput under given outage probability constraints in, both, the primary and secondary networks.

A new type of wireless networks, namely WPCNs, has been studied recently in \cite{16,17,18,19,20}. In WPCNs, wireless devices use the harvested RF energy to communicate with each other. WPCNs have been studied under various network setups; the wireless powered cellular network was investigated in \cite{16}, where power beacons are deployed randomly to charge the mobile devices. On the other hand, wireless powered sensor networks were investigated in \cite{17,18}, where a mobile charging vehicle is moving around in order to continuously provide sensor nodes with wireless energy. Moreover, \cite{19} proposed a new routing metric for wireless powered sensor networks based on the charging ability of the sensor nodes. In addition, the optimal charging and transmission cycles, with the objective of enhancing the lifetime of the network under user-specified end-to-end constraints (throughput and latency), have been characterized. Motivated by the fact that wireless energy transfer directly impacts data communication, since they both share the same frequency band, \cite{20} has proposed a distributed medium access protocol for efficiently sharing the radio resources for these two major functions.

An alternative model for WPCNs has recently attracted considerable attention in the literature \cite{21,22,23,24,25,26,27}. In this particular model, users first harvest RF energy on the downlink from wireless energy signals broadcast by a basestation (BS) or hybrid access point (HAP). Afterwards, users transmit their information signals to the HAP on the uplink using the energy harvested in the downlink phase, e.g., using TDMA in \cite{21}. In addition, \cite{22} introduced user cooperation as a solution to the doubly near-far phenomenon that results in unfair rate allocation among users as observed in \cite{21}. Furthermore, a full-duplex WPCN scheme has been introduced in \cite{23}. Taking into consideration the energy causality constraints, of practical significance, \cite{24} has studied full-duplex WPCNs in which a user can only consume energy harvested before its allocated uplink time for data transmission. Cognitive radio WPCNs have been introduced in \cite{25}, where the WPCN shares the same spectrum, for both downlink wireless energy transfer and uplink data transmissions, with the primary wireless communication system. In addition, the authors proposed two models for spectrum sharing, namely, underlay and overlay based cognitive WPCN, depending on the type of available information for the cognitive WPCN about the primary wireless communication system. Motivated by the fact that the location of HAPs and wireless energy nodes (WENs) would have a significant impact on the WPCN performance, the optimal node placement has been investigated in \cite{26}. The network deployment cost was minimized via characterizing the minimum number of HAPs and WENs needed to achieve the performance requirements of wireless devices. WPCNs with two types of nodes, with and without RF energy harvesting capability, was introduced in \cite{27}.

 In this paper, we generalize conventional TDMA wireless networks to a new type of wireless networks coined g-WPCNs, where nodes are assumed to be equipped with RF energy harvesting circuitries along with constant energy supplies. The prime motivation for this work is twofold: i) quantify the performance gains attributed to RF energy harvesting, when available to conventional TDMA wireless networks studied before and ii) relax the strong assumption adopted widely in prior WPCNs studies, whereby the user devices are solely operated by the inherently limited RF energy harvesting with no other sources of energy. Due to the limited amount of RF energy and the modest efficiency of harvesting circuitries, we argue that RF harvesting would predominantly serve as a supplementary energy source. Our prime objective is to optimize the design of g-WPCNs and characterize the gains obtained by the assumption that nodes have RF energy harvesting capabilities along with the constant energy supplies, compared to conventional TDMA wireless networks (with only constant energy supplies, yet, no energy harvesting) and WPCNs with only RF energy harvesting nodes \cite{21}.

Our main contribution in this paper is multi-fold. First, we introduce a new, more realistic wireless network setting, coined g-WPCNs, in which all nodes are equipped with RF energy harvesting circuitries along with the constant energy supplies. To the best of the authors' knowledge, this is the first generalized WPCNs optimization framework in the open literature. Second, we formulate an optimization problem to maximize the sum throughput under the generalized problem setting. Furthermore, we show that the generalized optimization problem seemlessly reduces to two extreme special cases in the literature, namely, conventional TDMA wireless networks with no RF energy harvesting capability and standard WPCNs with only RF energy harvesting nodes.
Third, we introduce WPCNs with two types of nodes, with and without RF energy harvesting capability, and characterize its optimal resource allocation policy in closed form.
Fourth, motivated by the fairness problem known for the sum throughput maximization objective, we formulate a maxmin problem for the generalized system setting. Finally, we establish convexity for the formulated problems and solve efficiently for the optimal policy using standard techniques. Our numerical results show that the two extreme network settings, namely, WPCNs with only RF energy harvesting nodes and conventional TDMA no-harvesting wireless networks, are considered as lower bounds on the performance of the generalized problem setting in terms of the maximum sum throughput and maxmin throughput. Moreover, the results reveal valuable insights and throughput-fairness trade-offs unique to our new problem setting.

The rest of the paper is organized as follows. The system model is presented in Section~\ref{sec:sys}. In Section~\ref{sec:gen}, the sum throughput maximization problem for the generalized system model is formulated,  convexity is established and an efficient algorithm is proposed to solve it. Furthermore, we show that formulations for extreme scenarios studied earlier in the literature fall as special cases of  the generalized problem formulation proposed here. In Section~\ref{sec: WPCNs with two type of nodes}, the sum throughput maximization problem of WPCNs with two types of nodes; with and without RF energy harvesting capability, is formulated. Furthermore, the optimal resource allocation policy is characterized in closed form. The maxmin throughput optimization problem is formulated, convexity is established and solved efficiently in Section~\ref{sec:maxmin}. Numerical results are presented in Section~\ref{sec:num}. Finally, Section~\ref{sec:con} concludes the paper and points out potential directions for future research.

\begin{figure} 
\centering
\includegraphics[width=9cm,height= 6 cm]{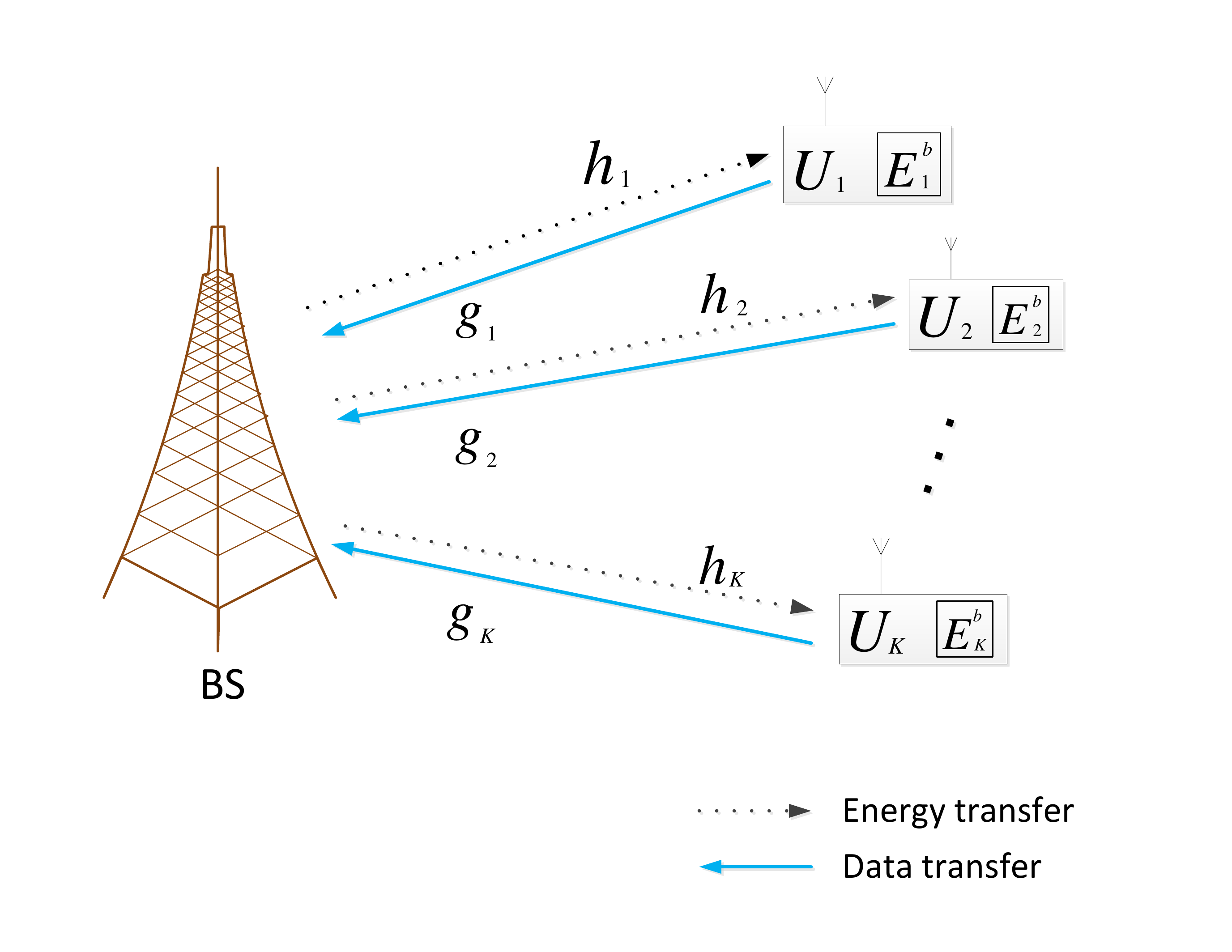}
    \caption{Generalized WPCN where nodes are powered with two energy sources.}
     \label{fig:1}
\end{figure}
%
%
%****************************************************************************************
\vspace{-0.5 cm}
\section{System Model}
\label{sec:sys}
\vspace{-0.2 cm}
We study a generalized wireless powered communication network consisting of one BS and $K$ users, as shown in Fig. \ref{fig:1}. It is assumed that the BS and all users are equipped with a single antenna each, operate over the same frequency channel and the radios are half-duplex. Each user, denoted by $U_{i}$ for $i=1,\cdots, K$, is assumed to be equipped with a constant energy supply, and thus has an allowable amount of energy to be consumed in each slot denoted by $E_{i}^{b}$ \cite{30,31,32}. Furthermore, each user is assumed to be equipped with an RF energy harvesting circuitry. In this paper, one of our main objectives is to characterize the performance gains attributed to having the RF energy harvesting capabilities, beyond conventional TDMA-based networks with no harvesting capabilities. 

The network operates in a TDMA fashion. For convenience, we assume the block (slot) duration is normalized to one. At the first $\tau_{0} \in [0,1]$ fraction of time, the BS broadcasts an energizing signal over the downlink so that each $U_{i}$ could harvest a certain amount of energy. The remaining $1-\tau_{0}$ fraction of time is allocated to uplink data transmissions where $U_{i}$ is assigned certain portion of time denoted by $\tau_{i}$\footnote{Note that slot time allocations are assumed to take continuous values. This, in turn, requires accurate synchronization methods to implement such scheme in realistic systems.}, for $i=1, \cdots, K$. Hence, the slot is split as follows.
\begin{equation} \label{eq1}
\sum_{i=0}^{K}{\tau_{i}} \leq 1.
\end{equation}

The downlink channel coefficient from the BS to $U_{i}$ and the uplink channel coefficient from $U_{i}$ to the BS are denoted by complex random variables $h_{i}^{\prime}$ and $g_{i}^{\prime}$, respectively, with channel power gains $h_{i} = \vert h_{i}^{\prime}\vert^{2}$ and $g_{i} = \vert g_{i}^{\prime} \vert^{2}$. It is assumed that all downlink and uplink channels are quasi-static flat fading, i.e., they remain constant over a time slot, but can change independently from one slot to another. The BS has perfect knowledge of the channel state information (CSI) to all users (i.e., all channel coefficients) at the beginning of each slot\footnote{The assumption that CSI is perfectly pre-estimated at the BS in the beginning of each slot is an idealization of actual practical systems. This calls for the necessity of using estimators with high accuracy to sufficiently reduce the potential estimation errors.}. The transmitted energy signal from the BS to all users, over the downlink, is denoted by $x_{B}$ with fixed average power, $P_B$, i.e., $E\left( \vert x_{B} \vert^{2}\right) = P_{B}$. Hence, the energy harvested by an arbitrary node, $U_{i}$, in the downlink phase is given by
\begin{equation}  \label{eq2}
E^h_{i} = \eta_{i} P_{B} h_{i} \tau_{0} ,
\end{equation}
where $\eta_{i}\footnote{Note that this paper falls within the context of WPCNs where the efficiency of energy harvesting circuitries is assumed to be linear \cite{21,22,23,24,25,26,27}. Incorporating the assumption of non-linear energy harvesting efficiency to our model is a challenging direction of future work.} \in (0,1)$ is the efficiency of the RF energy harvesting circuitry \cite{28,29}, at $U_{i}$. The value of $\eta_{i}$ depends on the efficiency of the harvesting antenna, the impedance matching circuit and the voltage multipliers. Therefore, the consumed energy per slot for uplink data transmission by $U_{i}$, $E_{i}$, is limited by
%\vspace{-0.5 cm}
\begin{equation} \label{eq3}
E_{i} \leq E^b_{i} + E^h_{i},\; i=1, \cdots, K.
\end{equation} 

\begin{table}[t!]\caption{Table of notation}
\centering
\begin{center}
\resizebox{0.5\textwidth}{!}{
    \begin{tabular}{ {c} | {c} }
    \hline\hline
    \textbf{Notation} & \textbf{Description} \\ \hline
    $E_{i}^{b}$; $E_{i}^{h}$ & Allowable amount of energy to be consumed by $U_{i}$ in each slot; amount of harvested energy by $U_{i}$  \\ \hline
    $\tau_{0}$; $\tau_i$ & Downlink energy transfer fraction of time; allocated portion of time to $U_{i}$ for uplink data transmissions\\ \hline
        $h_{i}$; $g_{i}$ & Downlink channel power gain from the BS to $U_{i}$; uplink channel power gain from $U_{i}$ to the BS\\ \hline
        $P_{B}$; $E_{i}; \sigma^2$ & Downlink energy transmit power by the BS; uplink consumed energy by $U_{i}$ for data transmission; noise power\\ \hline
        $E_{max}$ & Maximum allowable consumed energy by all users per slot\\ \hline
        $\tau_{1,i}$; $\tau_{2,j}$ & Uplink allocated time for $U_{1,i}$; uplink allocated time for $U_{2,j}$\\ \hline
        $\bar{E}$ & Amount of energy drawn by each $U_{2,j}$ from its dedicated energy supply within its assigned $\tau_{2,j}$\\ \hline
        $\eta_{i}$; $\beta$ & Efficiency of $U_{i}$'s RF energy harvesting circuitry; pathloss exponent\\ \hline
        $\Gamma$ & Signal to noise ratio gap due to a practical modulation and coding scheme used.\\ \hline\hline
    \end{tabular}}
\end{center}
\label{tab:TableOfNotations}
%\vspace{-8mm}
\end{table}

According to Shannon's formula, the achievable uplink throughput of $U_{i}$ in bits/second/Hz is given by
\begin{equation}  \label{eq5}
\begin{aligned}
R_{i} \left(E_{i},\tau_{i}\right) & = \tau_{i} \log_{2} \left(1 + \dfrac{g_{i} E_{i}}{\Gamma \sigma^{2} \tau_{i}}\right)\\
& =\tau_{i} \log_{2} \left(1 + \alpha_{i} \dfrac{E_{i}}{\tau_{i}}\right),
\end{aligned}
\end{equation}
where $\sigma^{2}$ is the noise power at the BS, $\alpha_{i} = \dfrac{g_{i}}{\Gamma \sigma^{2}}$ for $i=1, \cdots, K,$ and $\Gamma$ denotes the signal to noise ratio gap due to a practical modulation and coding scheme used. The notation used in this paper is summarized in Table~\ref{tab:TableOfNotations}.

%In this paper, we propose two optimal resource allocation schemes for WPCN systems with two types of nodes, namely, RF energy harvesting and legacy (battery-powered) nodes, which are different in the way Type II (legacy) nodes exploit their dedicated energy supplies. Under the first formulation, called Optimal Policy under per slot (Instantaneous) energy Constraint (OPIC), the consumed energy per slot by each legacy node ($E_{2,j}$) is optimized subject to maximum allowable energy consumption per slot, denoted $E_{max}$. Under the second formulation, called Optimal Policy under Average energy Constraint (OPAC), we relax the strong ``per slot energy requirement'' of OPIC. In OPAC, the energy consumption of Type II nodes is limited to a pre-specified fixed value per slot, denoted by $\bar{E}$, in an attempt to limit the overall system energy consumption. 
%
%
%***********************************************************************************************
%\vspace{-0.5 cm}
\section{Sum throughput maximization} \label{sec:gen}
%\vspace{-0.3 cm}
In this section, we formulate the sum throughput maximization problem for the generalized WPCN setting shown in Fig. 1 and establish its convexity which facilitates efficient solution using standard optimization solvers. We formulate the sum throughput maximization problem for a generalized setting of conventional TDMA-based wireless networks, whereby all nodes have RF energy harvesting capabilities along with the constant energy supplies. In particular, we find the optimal duration $\tau_0$ for harvesting as well as the durations, $\tau_{i}$, for uplink data transmissions and the optimal consumed energy by each user per slot, $E_{i}$, that maximize the system sum throughput subject to a system energy constraint \cite{27} on the total allowable consumed energy by all users per slot, denoted by $E_{max}$, the transmission slot duration constraint and the total allowable consumed energy by each user per slot constraints. The motivation behind introducing the system energy constraint is two-fold: i) it guarantees a fair comparison between our proposed g-WPCNs and other prior wireless networks, namely, conventional TDMA-based wireless networks (no RF energy harvesting) and WPCNs with RF energy harvesting nodes only, through setting $E_{max}$ with the average total amount of consumed energy in those prior wireless networks, and ii) it characterizes the maximum sum throughput that can be achieved by g-WPCNs via allocating the users that are closer to the BS, and hence experience better channels, more energy compared to other users, as will be highlighted in Section~\ref{sec:num}. Therefore, based on (\ref{eq1}) - (\ref{eq5}), the problem of maximizing the sum throughput per slot can be formulated as follows.
\begin{align}
\nonumber & \textbf{P1}: \hspace{0.5cm}  && \nonumber\underset{\mathbf{E},\pmb{\tau}}{\text{max}}\;\;  \sum_{i=1}^{K}{\tau_{i} \log_{2} \left(1 + \alpha_{i} \dfrac{E_{i}}{\tau_{i}}\right)} \\
\label{eq6a}&\text{s.t.} &&\sum_{i=1}^{K}{E_{i}} \leq E_{max},  \\
\label{eq6b} &&&\sum_{i=0}^{K}{\tau_{i}} \leq 1, \\
\label{eq6c} &&&  \pmb{\tau} \succeq \mathbf{0}, \\
\label{eq6d}&&& 0 \leq E_{i} \leq E^b_{i} + \eta_{i} P_{B} h_{i} \tau_{0},\hspace{0.5 cm} i=1, \cdots,K,
\end{align}
%\begin{equation*}  \label{eq6}
%\begin{aligned}
%& \textbf{P1}: & & \\  & \underset{\mathbf{E},\mathbf{\tau}}{\text{max}}&&  \sum_{i=1}^{K}{\tau_{i} \log_{2} \left(1 + \alpha_{i} \dfrac{E_{i}}{\tau_{i}}\right)} \\
%&\text{s.t.} &&\sum_{i=0}^{K}{\tau_{i}} \leq 1, \\
%&&&\sum_{i=1}^{K}{E_{i}} \leq E_{max},  \\
%&&&  \tau_{i} \geq 0, i=0, \cdots,K, \\
%&&& 0 \leq E_{i} \leq E^b_{i} + \eta_{i} P_{B} h_{i} \tau_{0}, i=1, \cdots,K, 
%\end{aligned}
%\end{equation*}
where $\pmb{\tau}=[\tau_{0}, \cdots,\tau_{K}]$, $\mathbf{E}=[E_{1}, \cdots, E_{K}]$, $\mathbf{0}$ is a vector of zeros that has the same size as $\pmb{\tau}$ and the symbol $\succeq$ represents the element-wise inequality.\\
\vspace{-0.4 cm}
\begin{theorem}\label{th:1} 
\textbf{P1} is a convex optimization problem.
\end{theorem}     
\begin{IEEEproof}
Please refer to Appendix A.
\end{IEEEproof}

 Based on Theorem~\ref{th:1}, \textbf{P1} is a convex optimization problem and, hence, can be solved efficiently using standard convex optimization solvers. Furthermore, it can be easily shown that there exists a $[\mathbf{E}\;\pmb{\tau}]$ policy that strictly satisfies all constraints of $\textbf{P1}$. Hence, according to Slater's condition \cite{35}, strong duality holds for this problem; therefore, the Karush-Kuhn-Tucker (KKT) conditions are necessary and sufficient for the global optimality of $\textbf{P1}$. However, due to the complexity of the problem, there are no closed form expressions that solve the KKT conditions. Therefore, in order to gain more insights about the optimal policy, we propose an algorithm based on alternating optimization approach for solving \textbf{P1}. First, we investigate the optimal time allocations ($\pmb{\tau}^{*}$) for a given $\mathbf{E}$ that satisfies (\ref{eq6a}) and $0 \leq E_{i} < E^b_{i} + \eta_{i} P_{B} h_{i}, i=1, \cdots,K$. Next, we get the optimal consumed energy allocations ($\mathbf{E}^{*}$) for a given $\pmb{\tau}$ that satisfies (\ref{eq6b}) - (\ref{eq6d}). Finally, the optimal time and energy allocations for \textbf{P1} are obtained by employing the alternating optimization procedure, as established by the following two Theorems and Algorithm 1.

\begin{theorem}\label{th:4} 
Given $\mathbf{E}$ that satisfies (\ref{eq6a}) and $0 \leq E_{i} < E^b_{i} + \eta_{i} P_{B} h_{i}, i=1, \cdots,K$, the optimal time allocations are given by
\begin{equation} \label{eq66}
\tau_{0}^{*} = \text{min} \left[ \left(\underset{i}{\text{max}}\lbrace\dfrac{E_{i} - E^b_{i}}{\eta_{i} P_{B} h_{i}}\rbrace\right)^{+},\; 1 \right],
\end{equation}
\begin{equation} \label{eq67}
\tau_{i}^{*}=\dfrac{\alpha_{i} E_{i} \left(1 - \tau_{0}^{*}\right)}{\sum_{j=1}^{K}{\alpha_{j} E_{j}}},\; i= 1, \cdots,K,
\end{equation}
where $(x)^{+} = \text{max}(0,x)$.
\end{theorem}     
\begin{IEEEproof}
Please refer to Appendix B.
\end{IEEEproof}

\begin{theorem}\label{th:5} 
Given $\pmb{\tau}$ that satisfies (\ref{eq6b}) - (\ref{eq6d}), the optimal energy allocations are given by
\begin{equation} \label{eq68}
E_{i}^{\ast} = 
  \begin{cases}
%  \begin{aligned}
  E^b_{i} + \eta_{i} P_{B} h_{i} \tau_{0},\;\text{if} \;E_{max} \geq E_{tot}\\
 \text{min}\left[\left(- \dfrac{\tau_{i}}{\alpha_{i}}\left(\dfrac{\alpha_{i}}{\lambda^{*} \ln(2)} + 1\right)\right)^{+},\;E^{b}_{i} + \eta_{i} P_{B} h_{i} \tau_{0} \right],\;\text{otherwise}
% \end{aligned}
 \end{cases}
\end{equation}
for $i = 1,\cdots, K$, where $E_{tot} = \sum_{j = 1}^{K}{\left(E^b_{j} + \eta_{j} P_{B} h_{j} \tau_{0}\right)}$ is the total amount of energy available for all users to be consumed per slot, and $\lambda^{*}$ satisfies the equality constraint $\sum_{i=1}^{K}{E_{i}^{*}} = E_{max}$.
\end{theorem}     
\begin{IEEEproof}
Please refer to Appendix C.
\end{IEEEproof}

\begin{algorithm}[h]
\caption{\textbf{P1} solver.}\label{euclid}
\begin{algorithmic}
 \STATE 1. Initialize: $t = 0$, $\mathbf{E} = \mathbf{E}^{(t)}$.
 \STATE 2. Repeat
 \STATE \hspace{1cm} (1) Compute $\pmb{\tau}^{(t+1)}$ from (\ref{eq66}) and (\ref{eq67}) with given $\mathbf{E}^{(t)}$ .
 \STATE \hspace{1cm} (2) Compute $\mathbf{E}^{(t+1)}$ from (\ref{eq68}) with given $\pmb{\tau}^{(t+1)}$.
 \STATE 3. Until $[\pmb{\tau}^{(t+1)}\; \pmb{E}^{(t+1)}]$ converges to a predetermined accuracy.
 \STATE 4. Set  $\pmb{\tau}^{*} = \pmb{\tau}^{(t+1)}$ and $\mathbf{E}^{*} = \mathbf{E}^{(t+1)}$.
\end{algorithmic}
\end{algorithm}

 According to Theorem~\ref{th:4} and Theorem~\ref{th:5}, for initial energy allocations ($\mathbf{E}^{(0)}$), the optimal time allocations $\pmb{\tau}^{(1)}$ can be obtained by (\ref{eq66}) and (\ref{eq67}). Afterwards, $\pmb{\tau}^{(1)}$ can be used to obtain $\mathbf{E}^{(1)}$ from (\ref{eq68}), and so on until $[\pmb{\tau}^{(t+1)}\; \pmb{E}^{(t+1)}]$ converges to a predetermined accuracy. Therefore, $\pmb{\tau}^{(t+1)}$ and $\mathbf{E}^{(t+1)}$ will be the optimal time and energy allocations for \textbf{P1}, respectively. The proposed alternating optimization approach is guaranteed to converge to the optimal solution of \textbf{P1} \cite{boyd2011alternating} since the objective function of \textbf{P1} is: 1) a concave function jointly in $\pmb{\tau}$ and $\mathbf{E}$ and 2) a smooth function in both $\pmb{\tau}$ and $\mathbf{E}$. At each iteration of Algorithm 1, the computational complexity of step 2.(1) is $\mathcal{O}(K+1)$ \cite{23} to obtain $\pmb{\tau}$ using (\ref{eq66}) and (\ref{eq67}). Furthermore, in step 2.(2), $\mathcal{O}(K)$ computations are required for computing $\mathbf{E}$ using (\ref{eq68}). Therefore, the complexity of one iteration of  Algorithm 1 is $\mathcal{O}(K+1)$, i.e., linear in the number of users.

 Next, we demonstrate the generality of \textbf{P1} through characterizing the conditions under which the sum throughput maximization problem for extreme scenarios known in the literature become special cases of our generalized formulation, namely, conventional TDMA-based wireless networks (no RF energy harvesting) and WPCNs with RF energy harvesting nodes only.
%
%*******************************************************************************************
%\vspace{-0.5cm}
\subsection{Prior formulations as special cases of \textbf{P1}} \label{sec:special}
A salient feature of the problem formulation in \textbf{P1} is its generality manifested through capturing the fact that wireless nodes in envisioned WPCNs are typically powered using multiple energy sources, namely, two sources (constant energy supplies and RF energy harvesting circuitries). This, in turn, gives rise to the key observation that related prior work would fall as special cases of \textbf{P1}. In this section, we present two conventional scenarios studied earlier in the literature as special cases of \textbf{P1} and introduce a third, more practical, special case in Section~\ref{sec: WPCNs with two type of nodes}.  

\subsubsection{Conventional TDMA-based wireless networks (no RF energy harvesting)}
In this scenario, all wireless nodes are legacy and, hence, are not equipped with RF energy harvesting circuitries, $(\tau^*_0=0)$, yet, have constant energy supplies. Hence, each user has an allowable amount of energy to be consumed in each slot, $E_{i}^{b}$ \cite{30,31,32}. Therefore, \textbf{P1} will reduce to the sum throughput maximization problem in conventional TDMA-based wireless networks as follows.   
\begin{equation}  \label{eq7}
\begin{aligned}
& \textbf{P2}: \hspace{0.5 cm}  &&\underset{\mathbf{E},\pmb{\tau^{\prime}}}{\text{max}}\;\; \sum_{i=1}^{K}{\tau_{i} \log_{2} \left(1 + \alpha_{i} \dfrac{E_{i}}{\tau_{i}}\right)} \\
&\text{s.t.} &&\sum_{i=1}^{K}{\tau_{i}} \leq 1, \\
&&&\sum_{i=1}^{K}{E_{i}} \leq E_{max},  \\
&&&  \pmb{\tau^{\prime}} \succeq 0,\\
&&&  0 \leq E_{i} \leq E^b_{i},\hspace{0.5 cm} i=1, \cdots,K,
\end{aligned}
\end{equation}
where $\pmb{\tau^{\prime}}=[\tau_{1}, \cdots,\tau_{K}]$.
Based on Theorem \ref{th:1}, $\textbf{P2}$ is a convex optimization problem, and thus can be solved using standard convex optimization techniques. Following the proof of Theorem \ref{th:4} and Theorem~\ref{th:5}, Algorithm 1 can solve $\textbf{P2}$ using the following expressions
\begin{equation} 
\tau_{i}^{*}=\dfrac{\alpha_{i} E_{i}}{\sum_{j=1}^{K}{\alpha_{j} E_{j}}},\; i= 1, \cdots,K,
\end{equation}
\begin{equation}
E_{i}^{\ast} = 
  \begin{cases}
   E^b_{i},\; \text{if} \;E_{max} \geq \bar{E}_{tot}\\
  \text{min}\left[\left(- \dfrac{\tau_{i}}{\alpha_{i}}\left(\dfrac{\alpha_{i}}{\lambda^{*} \ln(2)} + 1\right)\right)^{+},\;E^{b}_{i} \right],\; \text{otherwise}
 \end{cases}
\end{equation}
for $i = 1,\cdots, K$, where $\bar{E}_{tot} = \sum_{j = 1}^{K}{E^b_{j}}$ is the total amount of energy available for all users to be consumed per slot, and $\lambda^{*}$ satisfies the equality constraint $\sum_{i=1}^{K}{E_{i}^{*}} = E_{max}$.
\subsubsection{WPCNs with RF energy harvesting nodes only}
 According to the setting of \cite{21}, all nodes have RF energy harvesting capability only with no constant energy supplies, i.e., this implies that $E^b_{i} = 0$ in \textbf{P1}. Furthermore, all harvested energy by a user in a slot is fully consumed for uplink data transmission in the same slot, $E_{i}^{*} = \eta_{i} P_{B} h_{i} \tau_{0}$. In addition, there is no limitation on the allowable consumed energy per slot, i.e., staring from \textbf{P1}, we have that $E_{max} = \infty$. Therefore, \textbf{P1} reduces to the optimal time allocation problem maximizing the sum throughput in WPCNs with RF energy harvesting nodes only in \cite{21} as follows.
 \begin{equation}  \label{eq10}
\begin{aligned}
& \textbf{P3}: \hspace{0.5 cm}  && \underset{\pmb{\tau}}{\text{max}}\;\; \sum_{i=1}^{K}{\tau_{i} \log_{2} \left(1 + \gamma_{i} \dfrac{\tau_{0}} {\tau_{i}}\right)} \\
& \text{s.t.} & & \sum_{i=0}^{K}{\tau_{i}} \leq 1, \\
&&&  \pmb{\tau} \succeq \mathbf{0}, \\
\end{aligned}
\end{equation}
where $\gamma_{i} = \dfrac{\eta_{i} h_{i} g_{i} P_{B}}{\Gamma \sigma^{2}}$. The optimal time allocations of \textbf{P3} are given, according to \cite{21}, by
 \begin{equation} \label{eq11}
\tau_{i}^{\ast} = 
  \begin{cases}
  \dfrac{x^{*} - 1}{A + x^{*} - 1},\; \;  i = 0 \\
 \dfrac{\gamma_{i}}{A + x^{*} - 1},\; \; i = 1,\cdots,K,
 \end{cases}
\end{equation}
where $A = \sum_{i = 1}^{K}{\gamma_{i}}$ and $x^{*} > 0$ is the solution of $x\ln{x} - x + 1 = A$.
 
 It is obvious by now that the problem formulation \textbf{P1} is, indeed, a generalized formulation that encompasses two well-known problem settings in the literature, namely, conventional TDMA with no RF energy harvesting capability at the nodes and WPCNs with all nodes having solely RF energy harvesting capability only. Furthermore, we show in the next section that \textbf{P1} extends to cover an important scenario of practical significance, introduced in \cite{27} with two types of nodes, namely, RF energy harvesting nodes and legacy (no RF energy harvesting capability) nodes. 
%Such that RF harvesting nodes are assumed to have zero energy at the beginning of each slot and all the harvested %energy by each user per slot is used for uplink data transmission. While, the legacy nodes (no-harvesting) are assumed to %be equipped with dedicated unlimited energy supplies. WPCNs with two types of nodes will be presented in detail in the %next section.
%
 % 
\begin{figure} 
\centering
\includegraphics[width=9 cm,height= 6 cm]{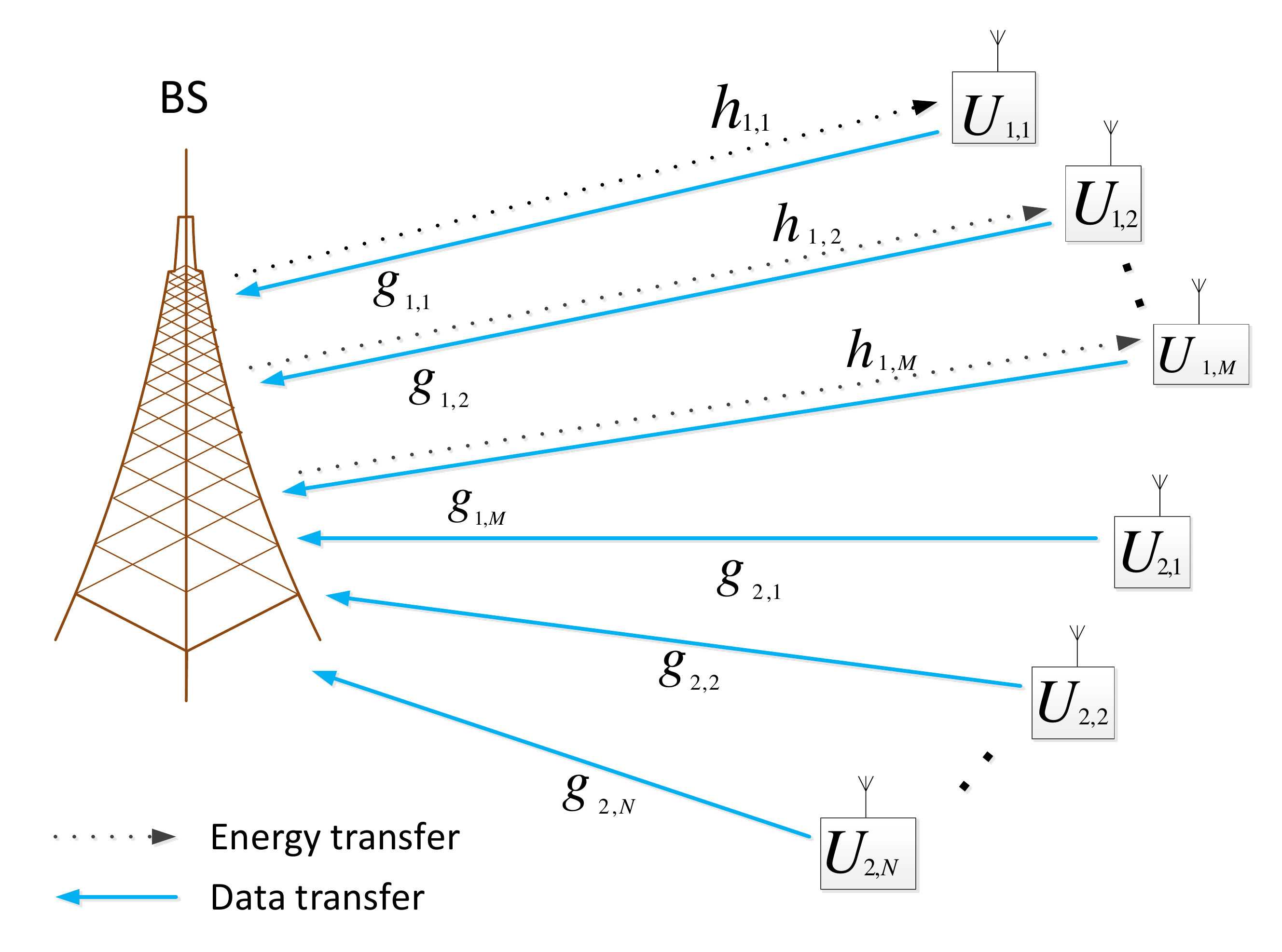}
    \caption{WPCN with heterogeneous nodes.}
     \label{fig:2}
\end{figure}
%
%***********************************************************************************************
\section{WPCNs with heterogeneous nodes} \label{sec: WPCNs with two type of nodes}

Motivated by the fact that RF energy harvesting is a new technology that may not be available to all the nodes in the network, we study in this section a practically viable network setting, namely, WPCNs with heterogeneous nodes. This constitutes an important step towards studying more realistic WPCNs since the RF energy harvesting technology would take time as it gradually penetrates the wireless industry. As shown in Fig. \ref{fig:2}, the network consists of two types of nodes; one is assumed to have RF energy harvesting capability and no other energy sources (Type I), denoted by $U_{1,i}$ for $i=1,\cdots,M$, while the other group has legacy nodes that are assumed not to have RF energy harvesting capability and are equipped with continuous energy supplies (Type II), denoted by $U_{2,j}$ for $j=1,\cdots,N$. Following the WPCNs operational regime, the BS with fixed power $(P_{B})$ broadcasts an energizing signal in the downlink over $\tau_{0}$ fraction of time. Afterwards, $U_{1,i}$ and $U_{2,j}$ are allocated portions of times for uplink data transmission, denoted by $\tau_{1,i}$ and $\tau_{2,j}$, respectively. It then follows that
\begin{equation} \label{eq12}
\tau_{0}+\sum_{i=1}^{M}{\tau_{1,i}}+\sum_{j=1}^{N}{\tau_{2,j}} \leq 1.
\end{equation}
 
 The downlink channel power gain from the BS to $U_{1,i}$, the uplink channel power gain from $U_{1,i}$ to the BS and the uplink channel power gain from $U_{2,j}$ to the BS are denoted by $h_{1,i}$, $g_{1,i}$ and $g_{2,j}$, respectively. Therefore, the achievable uplink throughput of $U_{1,i}$ and $U_{2,j}$ in bits/second/Hz is given by
\begin{equation}  \label{eq13}
\begin{aligned}
R_{1,i} \left(\tau_{0},\tau_{1,i}\right) & = \tau_{1,i} \log_{2} \left(1 + \dfrac{\eta_{i} P_{B} h_{1,i} g_{1,i} \tau_{0}}{\Gamma \sigma^{2}\tau_{1,i}}\right)\\
& =\tau_{1,i} \log_{2} \left(1 + \gamma_{i} \dfrac{\tau_{0}}{\tau_{1,i}}\right),
\end{aligned}
\end{equation}
\begin{equation}  \label{eq14}
\begin{aligned}
R_{2,j} \left(\bar{E},\tau_{2,j}\right) & = \tau_{2,j} \log_{2} \left(1 + \dfrac{g_{2,j} \bar{E}}{\Gamma \sigma^{2}\tau_{2,j}}\right) \\ 
& =\tau_{2,j} \log_{2} \left(1 + \theta_{j} \dfrac{\bar{E}}{\tau_{2,j}}\right),
\end{aligned}
\end{equation}
respectively, where $\bar{E}$ is the energy drawn by each $U_{2,j}$ from its dedicated energy supply within its assigned $\tau_{2,j}$ fraction of time, $\gamma_{i} = \dfrac{\eta_{i} h_{1,i} g_{1,i} P_{B}}{\Gamma \sigma^{2}}$ and $\theta_{j} = \dfrac{g_{2,j}}{\Gamma \sigma^{2}}$ for $i=1, \cdots, M$, $j=1, \cdots, N$. Therefore, from (\ref{eq13}) and (\ref{eq14}), the generalized formulation \textbf{P1} reduces to
\begin{equation}  \label{eq15}
\begin{aligned}
& \textbf{P4}: \; &&\underset{\pmb{\tau^{\prime  \prime}},\bar{E}}{\text{max}} \;\; \sum_{i=1}^{M}{R_{1,i} \left(\tau_{0},\tau_{1,i}\right)} + \sum_{j=1}^{N}{R_{2,j} \left(\bar{E},\tau_{2,j}\right)} \\
& \text{s.t.} & & \tau_{0}+\sum_{i=1}^{M}{\tau_{1,i}}+\sum_{j=1}^{N}{\tau_{2,j}} \leq 1, \\
&&& a\tau_{0} + N \bar{E} \leq E_{max},  \\
&&&  \pmb{\tau^{\prime  \prime}} \succeq \mathbf{0}, \\
&&& \bar{E} \geq 0, 
\end{aligned}
\end{equation}
where $\pmb{\tau^{\prime  \prime}}=[\tau_{0}, \tau_{1,1}, \cdots,\tau_{1,M}, \tau_{2,1}, \cdots,\tau_{2,N}]$ and $a = \sum_{i=1}^{M}{\eta_{i} P_{B} h_{1,i}}$.\\

In the following theorem, we characterize the optimal solution for \textbf{P4} in closed form which is one of the main contributions subject to this paper. 
\begin{theorem} \label{th:3} 
For $E_{max} > 0$, the optimal time and energy allocations of \textbf{P4} are given by (\ref{eq16}) - (\ref{eq19})
\begin{figure*}
   \begin{equation}  \label{eq16}
  \tau_{0}^{\ast} = \begin{cases}
                      \begin{aligned}
  &\min \left[\dfrac{x^{*} - 1}{A_{1} + x^{*} - 1} , \dfrac{E_{max}}{a}\right],&& \text{if} \; E_{max} \leq \dfrac{a(x_{1}^{*} - 1)}{A_{1} + x_{1}^{*} - 1} \; \text{and} \;A_{1} \geq\dfrac{a}{N} A_{2} \\
                           &\dfrac{N\left(x_{1}^{\ast} - 1 \right)- E_{max} A_{2}  }{N\left(x_{1}^{\ast} - 1 + A_{1}\right) - a A_{2} },&& \text{if} \; \dfrac{a(x_{1}^{*} - 1)}{A_{1} + x_{1}^{*} - 1} \leq E_{max} \leq \dfrac{N}{A_{2}}(x_{1}^{*} - 1) \; \text{and} \;A_{1} \geq \dfrac{a}{N} A_{2}\\
                                   &0 ,&& \text{if} \; \left( E_{max} \geq \dfrac{N}{A_{2}}(x_{1}^{*} - 1)\; \text{and} \;A_{1} \geq\dfrac{a}{N} A_{2} \right) \\ &&& \hspace{2 cm}\text{or}  \;\left( A_{1} < \dfrac{a}{N} A_{2}\right )
                      \end{aligned}
                  \end{cases}
\end{equation}
 \begin{equation}  \label{eq17}
 \tau_{1,i}^{\ast} =
 \begin{cases}
   \begin{aligned}
&\max \left[\dfrac{\gamma_{i}}{A_{1} + x^{*} - 1} , \dfrac{\gamma_{i}}{A_{1}}\left(1 - \dfrac{E_{max}}{a}\right)\right],&&\text{if} \; E_{max} \leq \dfrac{a(x_{1}^{*} - 1)}{A_{1} + x_{1}^{*} - 1} \; \text{and} \;A_{1} \geq\dfrac{a}{N} A_{2} \\
  &\dfrac{\gamma_{i} \left(N\left(x_{1}^{\ast}- 1\right) - E_{max} A_{2} \right) }{\left(x_{1}^{\ast} - 1\right) \left(N\left(x_{1}^{\ast} - 1 + A_{1}\right) - a A_{2} \right)},&&\text{if} \; \dfrac{a(x_{1}^{*} - 1)}{A_{1} + x_{1}^{*} - 1} \leq E_{max} \leq \dfrac{N}{A_{2}}(x_{1}^{*} - 1) \\ &&& \hspace{2 cm}\text{and} \;A_{1} \geq \dfrac{a}{N} A_{2} \\
  &0,&&\text{if} \; \left(E_{max} \geq \dfrac{N}{A_{2}}(x_{1}^{*} - 1)\; \text{and} \;A_{1} \geq\dfrac{a}{N} A_{2}\right)\\ &&& \hspace{2 cm}\text{or}\left(A_{1} < \dfrac{a}{N} A_{2}\right) 
 \end{aligned}
 \end{cases}
 \end{equation}
 \begin{equation}  \label{eq18}
 \tau_{2,j}^{\ast} =
 \begin{cases}
 \begin{aligned}
  &0,&&\text{if} \; E_{max} \leq \dfrac{a(x_{1}^{*} - 1)}{A_{1} + x_{1}^{*} - 1} \; \text{and} \;A_{1} \geq\dfrac{a}{N} A_{2} \\
  &\dfrac{\theta_{j} \left(E_{max}\left(x_{1}^{\ast} - 1 + A_{1} \right) - a \left(x_{1}^{\ast} - 1\right)\right) }{\left(x_{1}^{\ast} - 1\right)\left(N\left(x_{1}^{\ast} - 1 + A_{1}\right) - a A_{2} \right)}, &&\text{if} \; \dfrac{a(x_{1}^{*} - 1)}{A_{1} + x_{1}^{*} - 1} \leq E_{max} \leq \dfrac{N}{A_{2}}(x_{1}^{*} - 1) \\ &&& \hspace{2 cm}\text{and} \;A_{1} \geq \dfrac{a}{N} A_{2} \\
   &\dfrac{\theta_{j}}{A_{2}},&&\text{if} \; \left(E_{max} \geq \dfrac{N}{A_{2}}(x_{1}^{*} - 1)\; \text{and} \;A_{1} \geq\dfrac{a}{N} A_{2}\right)\\ &&& \hspace{2 cm}\text{or}\left(A_{1} < \dfrac{a}{N} A_{2}\right) 
   \end{aligned}
 \end{cases}
 \end{equation}
 \begin{equation}  \label{eq19}
 \bar{E}^{\ast} =
 \begin{cases}
  \begin{aligned}
  &0,&&\text{if} \; E_{max} \leq \dfrac{a(x_{1}^{*} - 1)}{A_{1} + x_{1}^{*} - 1} \; \text{and} \;A_{1} \geq\dfrac{a}{N} A_{2} \\
  &\dfrac{E_{max}\left(x_{1}^{\ast} - 1 + A_{1} \right) - a \left(x_{1}^{\ast} - 1\right) }{N\left(x_{1}^{\ast} - 1 + A_{1}\right) - a A_{2}},&&\text{if} \; \dfrac{a(x_{1}^{*} - 1)}{A_{1} + x_{1}^{*} - 1} \leq E_{max} \leq \dfrac{N}{A_{2}}(x_{1}^{*} - 1) \\ &&& \hspace{2 cm}\text{and} \;A_{1} \geq \dfrac{a}{N} A_{2} \\
   &\dfrac{E_{max}}{N},&&\text{if} \; \left(E_{max} \geq \dfrac{N}{A_{2}}(x_{1}^{*} - 1)\; \text{and} \;A_{1} \geq\dfrac{a}{N} A_{2}\right)\\ &&& \hspace{2 cm}\text{or}  \;\left(A_{1} < \dfrac{a}{N} A_{2}\right) 
   \end{aligned}
 \end{cases}
 \end{equation}
 \hrulefill
\end{figure*}
 for $i=1,\cdots,M$ and $j=1,\cdots, N$, where $A_{1} = \sum_{i=1}^{M}{\gamma_{i}}$, $A_{2} = \sum_{j=1}^{N}{\theta_{j}}$, $x_{1}^{\ast} > 1$ is the solution of $f(x_{1}) = A_{1} - \dfrac{a}{N} A_{2}$ and $x^{\ast} > 1$ is the solution of $f(x) = A_{1}$, where
 \begin{equation}  \label{eq20}
 f(x) = x \ln(x) - x + 1.\\
 \end{equation}
 \end{theorem}
\begin{IEEEproof}
 Please refer to Appendix D.
\end{IEEEproof}

For the sake of obtaining more insight into the solution given in Theorem~\ref{th:3}, we consider next a simple WPCN with only two users; one user of each type mentioned before.
 \subsection*{A Two-User Example}
 With the objective of capturing the optimality criteria of $\textbf{P4}$, we study a simple WPCN of only two nodes where $M=1$ and $N=1$. Referring to Theorem~\ref{th:3}, 
few key observations about the optimal solution are now in oder. First, the energy harvesting node is only allocated portion of the slot duration (either for harvesting $\tau_{0}$ or for data transmission $\tau_{1,1}$) if its uplink channel power gain $(g_{1,1})$ is greater than the channel power gain of the legacy node $(g_{2,1})$. Otherwise, the whole slot and the total allowable energy consumption per slot $(E_{max})$ are assigned to the legacy node. Second, for $g_{1,1} \geq g_{2,1}$, the portion of time which is allocated to the energy harvesting node depends on $E_{max}$. Based on the value of the maximum system energy consumption allowed per slot, $(E_{max})$, three different cases arise as follows. For small $E_{max} \leq \dfrac{a(x_{1}^{*} - 1)}{\gamma_{1} + x_{1}^{*} - 1}$, the energy harvesting node is allocated the whole slot and consumes the entire $E_{max}$. On the other hand, for large $E_{max} \geq \dfrac{1}{\theta_{1}}(x_{1}^{*} - 1)$, the whole slot and $E_{max}$ are assigned to the legacy node, as intution suggests. Finally, for $\dfrac{a(x_{1}^{*} - 1)}{\gamma_{1} + x_{1}^{*} - 1} \leq E_{max} \leq \dfrac{1}{\theta_{1}}(x_{1}^{*} - 1)$, each user is assigned a slot portion for uplink data transmission which is proportional to its uplink channel power gain. Taking into consideration the above two observations, the sum throughput maximization problem causes unfairness to different users. In addition, we note that each node is allocated uplink transmission time which does not only depend on its uplink channel power gain, as in WPCNs with energy harvesting nodes only, but also depends on the amount of allowable energy consumption per slot.

\begin{figure}
   \centering
\includegraphics[width=9 cm, height= 6cm]{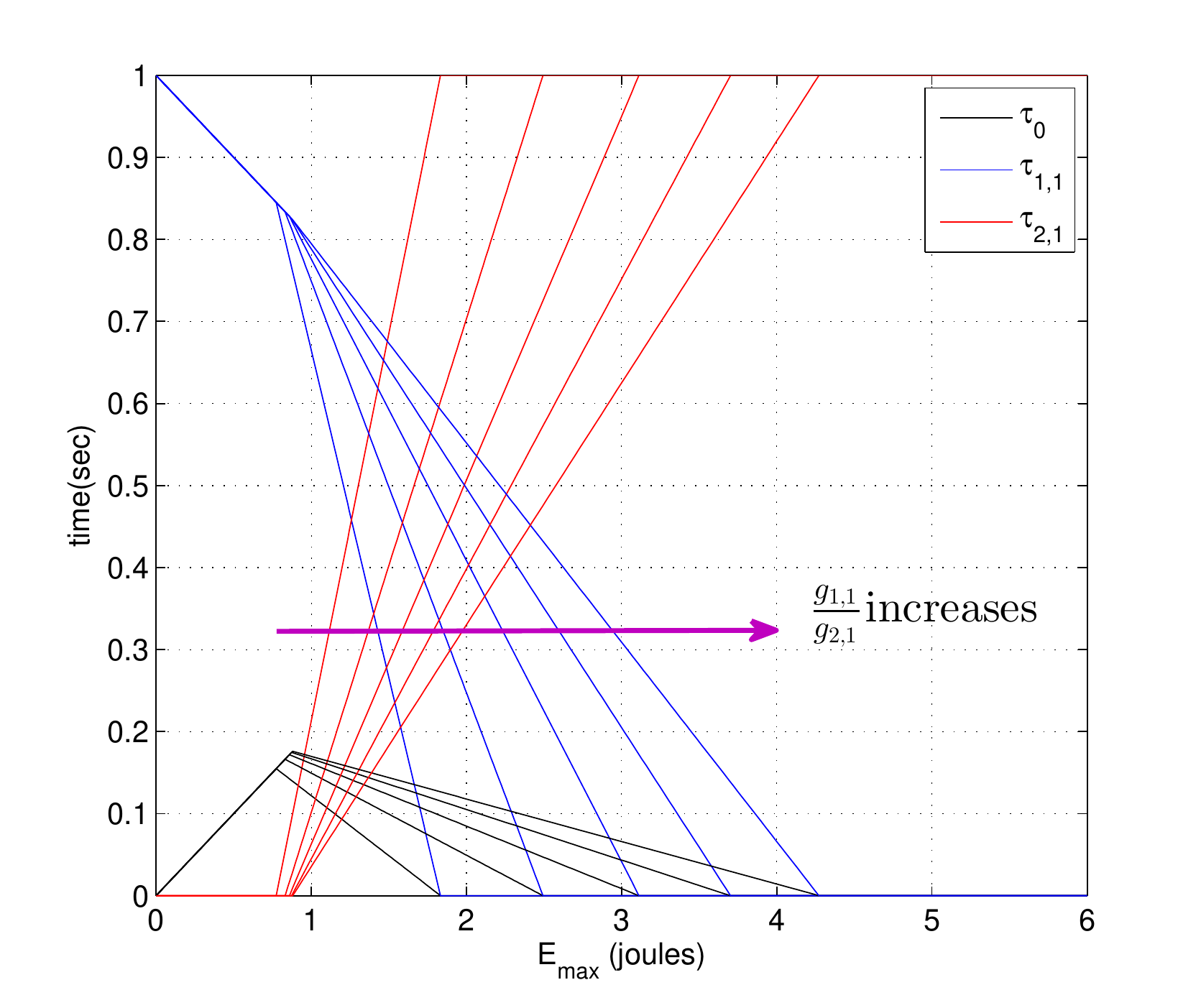}
    \caption{Optimal time allocation behavior with $E_{max}$ for a two user system; one user of each type.}
     \label{fig:3}
\end{figure}

Fig. \ref{fig:3} shows the optimal time allocation, for a WPCN with two nodes where $M=1$ and $N=1$ vs. $E_{max}$ for different values of $\dfrac{g_{1,1}}{g_{2,1}}$ ($\dfrac{g_{1,1}}{g_{2,1}} =$ 2, 2.5, 3, 3.5 and 4). We fix $a = 5$ and $\theta_{1} = 20$. It is observed that as $\dfrac{g_{1,1}}{g_{2,1}}$ expands, the range of $E_{max}$ values for which both users are allocated portions of the slot duration for data transmission, given by $\dfrac{a(x_{1}^{*} - 1)}{\gamma_{1} + x_{1}^{*} - 1} \leq E_{max} \leq \dfrac{1}{\theta_{1}}(x_{1}^{*} - 1)$, expands. It is also worth noting that $\tau_{2,1}$ monotonically increases and $\tau_{1,1}$ monotonically decreases as $E_{max}$ increases over the shown range.

%Based on Theorem ~\ref{th:2}, it is clear that the optimal time allocated to each user for uplink information transmission depends on its distance to the BS, i.e., the near users (with better channel power gains) to the BS are allocated more uplink transmission time than the far users, which demonstrates the doubly near-far phenomenon \cite{7}. Moreover, it is observed that $\tau_{2,j}^{\ast}$ is proportional to $E_{max}$, i.e., as the amount of the allowable consumed energy per slot $(E_{max})$ increases, the uplink allocated time for legacy nodes increases and that allocated for RF energy harvesting nodes decreases. Taking into consideration the above two observations, the sum throughput maximization results in an unfair achievable throughput among different users. 
%
It is obvious by now that the optimal resource allocation policy, which maximizes the sum throughput, in WPCNs with heterogeneous nodes depends on two major factors: 1) the total amount of allowable energy consumption per slot $(E_{max})$ and 2) the channel power gains of the different nodes. This, in turn, leads to unfair rate allocation among different users as shown above. In the next section, we propose to maximize the minimum throughput to tackle the fairness problem.
%
%
%***********************************************************************************************
\section{Fair rate allocation in generalized WPCNs} \label{sec:maxmin}
In this section, we shift our attention to the fair rate allocation problem in generalized WPCNs. This is motivated by the fairness challenges faced by \textbf{P1} as discussed next. In particular, we formulate a maxmin rate allocation problem.
%
%***************************************************************************
\subsection{Motivation}
\begin{figure}
   \centering
\includegraphics[width=9 cm, height= 5.5cm]{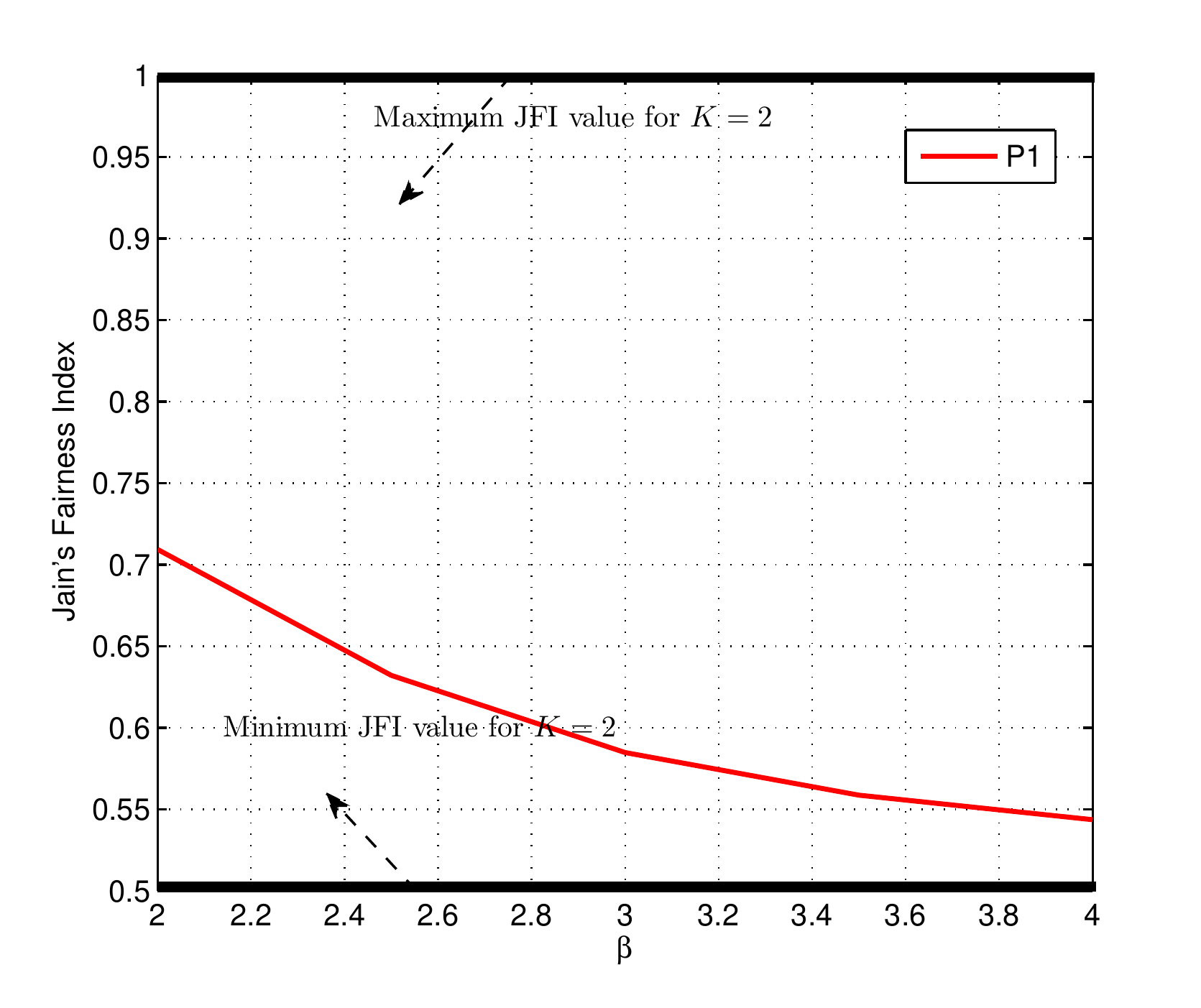}
    \caption{Jain's fairness index for the optimal solution of $\textbf{P1}$ vs. the pathloss exponent for $K=2$ users.}
     \label{fig:4}
\end{figure}

Given the sum throughput maximization problem in $\textbf{P1}$, the total allowable consumed energy per slot constraint in (\ref{eq6a}) allocates more energy, and, hence more uplink transmission time to nodes with better channel power gains. This leads to unfair rate allocation among different users. In Fig. \ref{fig:4}, the Jain's fairness index (JFI) \cite{33} is plotted for the optimal solution of \textbf{P1} against the pathloss exponent for a WPCN with two users, $K = 2$. Generally, JFI is defined as $\dfrac{\left( \sum_{i=1}^{K}{R_{i}}\right)^{2}}{K \sum_{i=1}^{K}{R_{i}^{2}}}$, where $R_i$ is the rate allocated to $U_{i}$. The channel power gains are modeled as $h_{i} = g_{i} = 10^{-3} \rho_{i}^{2} d_{i}^{-\beta}$ for $i=1, \cdots, K$, where $d_{i}$ denotes the distance between $U_{i}$ and the BS, $\beta$ denotes the pathloss exponent and $\rho_{i}$ is the standard Rayleigh short term fading; therefore $\rho_{i}^{2}$ is exponentially distributed random variable with unit mean. In addition, $P_{B} = 20$ dBm, $E^b_{1} = E^b_{2} = 10^{-7}$ joules, $\sigma^{2} = -160$ dBm/Hz, $\eta_{1} = \eta_{2} = 0.5$, $\Gamma = 9.8$ dB, $ d_{1} = \dfrac{d_{2}}{2} = 5$ meters, $E_{max} = 10^{-6} $ joules and the bandwidth is set to be 1 MHz. In addition, each throughput value is obtained by averaging over 1000 randomly generated channel realizations. For a wireless network of two users, the JFI ranges from 0.5 (worst case) to 1 (best case) and it is maximum when the two users achieve the same throughput. It is observed that the fairness index monotonically decreases as the pathloss exponent increases until it nearly approaches its worst value (0.5) when $\beta = 4$. This happens since the gap between the users' channel power gains increases as the pathloss exponent increases. This, in turn, highlights one instance of the fundamental throughput-fairness trade-off, where the maximum sum throughput is achieved at the expense of a modest degradation in the fairness.
%
%****************************************************************************
\subsection{Generalized Maxmin fairness formulation}
     Motivated by the fairness limitations of $\textbf{P1}$, we propose an alternative generalized optimization problem targeting fairness in the well-known maxmin sense \cite{34} subject to the same constraints of $\textbf{P1}$ as follows.
   \begin{equation}  \label{eq25}
 \begin{aligned}
& \textbf{P1}^{\text{Maxmin}}: \hspace{0.5 cm} &&\underset{\mathbf{E},\pmb{\tau}}{\text{max}} \; \; \underset{i}{\text{min}} \left(R_{i} \left(E_{i},\tau_{i}\right)\right)  \\
&\text{s.t.} &&\sum_{i=0}^{K}{\tau_{i}} \leq 1, \\
&&&\sum_{i=1}^{K}{E_{i}} \leq E_{max},  \\
&&&  \pmb{\tau} \succeq \mathbf{0},\\
&&& 0 \leq E_{i} \leq E^b_{i} + \eta_{i} P_{B} h_{i} \tau_{0},\hspace{0.5 cm} i=1, \cdots,K. 
\end{aligned}
\end{equation} 
Based on Theorem~\ref{th:1}, it follows that the objective function of problem $\textbf{P1}^{\text{Maxmin}}$ which is the minimum of a set of concave functions, i.e, $R_{i} \left(E_{i},\tau_{i}\right)$ for $i=1,\cdots,K$, is a concave function. Therefore, $\textbf{P1}^{\text{Maxmin}}$ is a convex optimization problem. Note that, for the same conditions discussed in sections~\ref{sec:gen} and~\ref{sec: WPCNs with two type of nodes}, under which \textbf{P1} reduces to the sum throughput maximization problem for extreme scenarios known in the literature, $\textbf{P1}^{\text{Maxmin}}$ also reduces to the maxmin problem in these extreme cases. An equivalent optimization problem to $\textbf{P1}^{\text{Maxmin}}$ can be cast as follows.
 \begin{equation}  \label{eq26}
 \begin{aligned}
& \textbf{P1}^{-\text{Maxmin}}: & & \underset{t,\mathbf{E},\pmb{\tau}}{\text{max}} \; t  \\
& \text{s.t.} & & \tau_{i} \log_{2} \left(1 + \alpha_{i} \dfrac{E_{i}}{\tau_{i}}\right) \geq t,\hspace{0.5 cm} i=1, \cdots,K, \\
&&&\sum_{i=1}^{K}{E_{i}} \leq E_{max},  \\
&&&  \pmb{\tau} \succeq \mathbf{0}, \\
&&& 0 \leq E_{i} \leq E^b_{i} + \eta_{i} P_{B} h_{i} \tau_{0}, \hspace{0.5 cm}i=1, \cdots,K, 
\end{aligned}
\end{equation} 
where $t$ is an auxiliary variable that denotes the minimum throughput achieved by each user.   

With the purpose of obtaining more insight into the optimal policy of $\textbf{P1}^{-\text{Maxmin}}$, we provide the following Theorem which shows that the optimal policy of $\textbf{P1}^{-\text{Maxmin}}$ must satisfy the condition that all users achieve the same throughput.
\begin{theorem}\label{th:6} 
The optimal policy of $\textbf{P1}^{-\text{Maxmin}}$ satisfies $R_i\left(E_{i}^*,\tau_{i}^*\right) = t^*$ for $i = 1,\cdots, K$.
\end{theorem}     
\begin{IEEEproof}
The proof is by contradiction. Without loss of generality, assume that the optimal policy satisfies $R_{i}\left(E_{i}^{*},\tau_{i}^{*}\right) = t_{1}$, $i = 1,\cdots, K-1$, and $R_{K}\left(E_{K}^*,\tau_{K}^*\right) = t_{2}$. Furthermore, assume that $t_{1} < t_{2}$, and, hence $t^{*} = t_{1}$. The monotonicity of each individual $R_{i}(E_{i},\tau_{i})$ in both $(E_{i},\tau_{i})$ guarantees that we can find $[\mathbf{E}^{\prime}\; \pmb{\tau}^{\prime}]$ which improves the minimum achievable throughput by all users. This can be achieved through decreasing $E_{k}^*$ or $\tau_{K}^{*}$ while increasing $E_{i}^{*}$ or $\tau_{i}^{*}$, $i = 1,\cdots, K-1$, till all users achieve a common throughput $t^{\prime}$ ($t_{1} < t^{\prime} < t_{2}$) and then no further improvements can be done. Therefore, the achievable throughput by all users using $[\mathbf{E}^{\prime}\; \pmb{\tau}^{\prime}]$ will be $t^{\prime} > t_{1}$ which contradicts with the assumption that $t_{1}$ is the maxmin throughput. This establishes the proof.
\end{IEEEproof}  
Due to the convexity of $\textbf{P1}^{-\text{Maxmin}}$ and based on Theorem~\ref{th:6}, $\textbf{P1}^{-\text{Maxmin}}$ could be solved efficiently using standard convex optimization techniques, e.g, the sub-gradient approach along with the alternating optimization procedure. Details are omitted due to space limitations. One subgradient approach based algorithm is proposed to solve the maxmin problem in WPCNs with RF energy harvesting nodes only in \cite{21}. In the next section, we compare the two generalized formulations with respect to the total system throughput and individual user's throughput in order to highlight the merits and limitations of both.
%
%
%***************************************************************************

\begin{table}[t!]\caption{Table of simulation parameters}
\centering
\begin{center}
\resizebox{0.5\textwidth}{!}{
    \begin{tabular}{ {c} | {c} }
    \hline\hline
    \textbf{Parameter} & \textbf{Value} \\ \hline
        $h_{i} = g_{i}$ & $10^{-3} \rho_{i}^{2} d_{i}^{-\beta}$ ($\rho_{i}^{2}$ is exponentially distributed random variable with unit mean)\\ \hline
        $d_{1} = d_{1,1}$; $d_{2} = d_{2,1}$ & $10$ meters; 5 meters\\ \hline
        $\sigma^2$; bandwidth & $-160$ dBm/Hz; $1$ MHz\\ \hline
        $\eta_{1} = \eta_{2}$; $\Gamma$ & $0.5$; $9.8$ dB\\ \hline\hline
    \end{tabular}}
\end{center}
\label{tab:TableOfsimulation}
%\vspace{-8mm}
\end{table}
\section{Numerical results}
\label{sec:num}
\subsection{System setup}
 We provide numerical results showing the merits of the formulated optimization problems and the associated trade-offs. Motivated by the convexity of the formulated maxmin problem in Section~\ref{sec:maxmin}, we use standard optimization solvers, e.g., CVX \cite{35}, to obtain its optimal solution. We denote the maxmin formulation of \textbf{Pi} by $\textbf{Pi}^{\text{Maxmin}}$, $i \in \{1,2,3,4\}$. We consider same parameters as in \cite{21} as follows. If not otherwise stated, we consider the following parameters $P_{B} = 30$ dBm, $\sigma^{2} = -160$ dBm/Hz, $\eta_{i}= 0.5$ for $i=1,\cdots,K$, $\Gamma = 9.8$ dB and the bandwidth is set to be 1 MHz. In addition, we consider the same model for the channel power gains as in Fig. \ref{fig:4}. Moreover, each throughput curve shown later is obtained by averaging over 1000 randomly generated channel realizations. In Fig. 5, 6, 7, 8 and 9, we consider the same scenario for all studied networks. The WPCN with two types of nodes is assumed to have $N=1$, $M=1$, $d_{1,1} = 10$ meters and $d_{2,1} = 5$ meters. In addition, the other wireless networks are considered to have two users with the same $d_{1,1}$ and $d_{2,1}$ given above. The average maximum sum throughput and maxmin throughput of the generalized problem setting (\textbf{P1} and $\textbf{P1}^{\text{Maxmin}}$) and conventional TDMA-based wireless networks (\textbf{P2} and $\textbf{P2}^{\text{Maxmin}}$) are plotted for different values of $E^b_{i}$ ($E^b_{1}$ = $E^b_{2}$ = $3\times10^{-7}$, $7\times10^{-7}$ and $5\times10^{-6}$ joules). The used values of simulation parameters are summarized in Table \ref{tab:TableOfsimulation}.
 
 Our objective it to fairly compare the performance of the generalized problem setting with the performance of different wireless networks discussed in sections~\ref{sec:gen} and ~\ref{sec: WPCNs with two type of nodes}, namely, conventional TDMA-based wireless networks, WPCNs with energy harvesting nodes only and WPCNs with two types of nodes subject to same amount of available resources. Towards this objective, for the sum throughput optimization problems, the average amount of harvested energy over the 1000 channel realizations for the WPCN with only energy harvesting nodes (\textbf{P3}) is set to $E_{max}$ in \textbf{P1}, \textbf{P2} and \textbf{P4} (i.e., per slot system energy constraint). This, in turn, results in the same long-term average energy consumption in all systems. Similarly, for the maxmin throughput optimization problems, the average amount of harvested energy over the 1000 channel realizations for the WPCN with only energy harvesting nodes ($\textbf{P3}^{\text{Maxmin}}$) is set to $E_{max}$ in $\textbf{P1}^{\text{Maxmin}}$, $\textbf{P2}^{\text{Maxmin}}$ and $\textbf{P4}^{\text{Maxmin}}$. Finally, in Fig. 10, we show the impact of replacing a number of Type I nodes with Type II nodes on WPCNs with heterogeneous nodes performance (\textbf{P4}).
 \begin{figure}[t!]
\centering
\includegraphics[width=9 cm, height= 7cm]{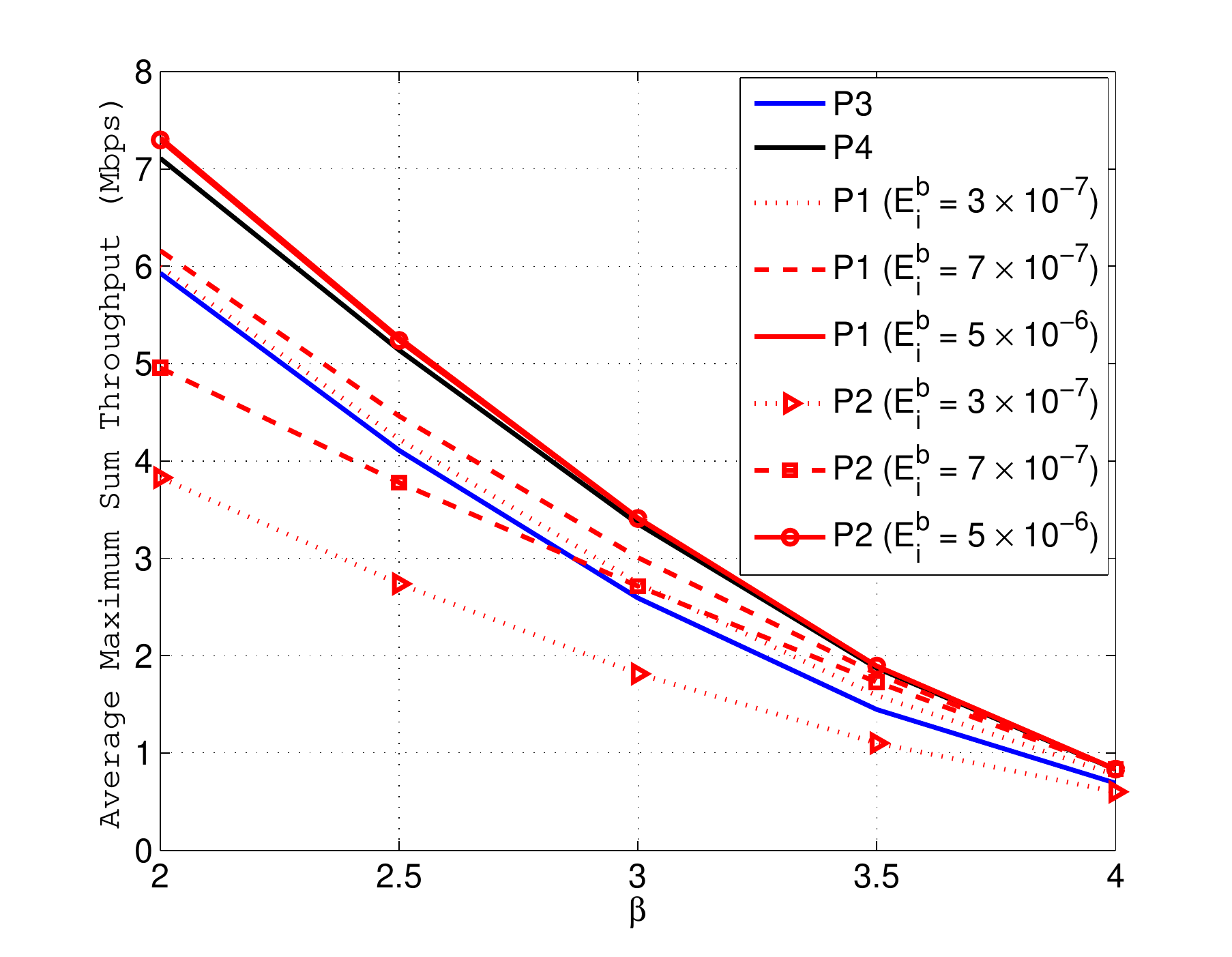}
    \caption{Average maximum sum throughput for all systems with two nodes vs. the Pathloss Exponent, $\beta$.}
     \label{fig:5}
\end{figure}
\subsection{Performance results}

%We consider the same scenario for all studied networks. The WPCN with two types of nodes is assumed to have $N=1$, $M=1$, $d_{1,1} = 10$ meters and $d_{2,1} = 5$ meters. In addition, the other wireless networks are considered to have two users with the same $d_{1,1}$ and $d_{2,1}$ given above. The average maximum sum throughput of the generalized problem setting (\textbf{P1}) and conventional TDMA-based wireless networks (\textbf{P2}) is plotted for different values of $E^b_{i}$ ($E^b_{1}$ = $E^b_{2}$ = $3\times10^{-7}$, $7\times10^{-7}$ and $5\times10^{-6}$ joules). Our objective is to fairly compare the four systems with the same total amount of energy resources. Towards this objective, at each pathloss exponent value, the average amount of harvested energy over the 1000 channel realizations for the WPCN with only energy harvesting nodes (\textbf{P3}) is set to $E_{max}$ in \textbf{P1}, \textbf{P2} and \textbf{P4} (i.e., per slot system energy constraint). This, in turn, results in the same long-term average energy consumption in all systems.

In Fig. \ref{fig:5}, we compare the maximum sum throughput, averaged over 1000 channel realizations, for the 4 studied systems vs. the pathloss exponent $(\beta)$. A number of observations are now in order. First, we note that the average maximum sum throughput of the four studied systems monotonically decreases as the pathloss exponent increases. This is due to the fact that the channel power gains become worse as $\beta$ increases. Therefore, the amount of harvested energy by each user becomes lower and, hence, the average maximum sum throughput decreases. Second, when $E_{i}^{b} = 3 \times 10^{-7} \; \text{and}\; 7 \times 10^{-7}$ Joules, the average maximum sum throughput attained by \textbf{P1} is notably larger than that of \textbf{P2} for $\beta \leq 3$. This, in turn, highlights the great influence of the RF energy harvesting capability on generalized WPCNs performance, compared to conventional TDMA wireless networks (no RF energy harvesting). More specifically, when $\beta \leq 3$, both users experience good channels, and thus the amount of harvested energy is so large that the performance of \textbf{P1} greatly outperform that of \textbf{P2}. On the other hand, as $\beta$ increases $(\beta > 3)$, the channel power gains become worse, and, hence the effectiveness of the RF energy harvesting capability on the network performance decreases. Therefore, the performance of \textbf{P1} approaches that of \textbf{P2}. Third, when $E^b_{i}$ is large, i.e., $E_{i}^b = 5 \times 10^{-6}$, both $\textbf{P1}$ and $\textbf{P2}$ achieve the same average maximum sum throughput. This happens since when $E_{i}^{b}$ is large, the average maximum sum throughput of \textbf{P1} is attained via allocating the entire slot duration for uplink data transmissions, i.e., $\tau_{0} = 0$. Finally, the average maximum sum throughput achieved by \textbf{P4} is higher than the average maximum sum throughput achieved by \textbf{P3} due to the fact that the total allowable energy consumption per slot constraint allocates more energy to the user with higher channel power gains, that is, the legacy node in our scenario, to maximize the sum throughput. Therefore, in our scenario, the average maximum sum throughput is attained via allocating more energy to the legacy node than the energy harvesting node and, hence, reducing $\tau_{0}$. This is to the contrary of the WPCN with energy harvesting nodes only (\textbf{P3}), where the amount of harvested energy by farther user cannot be efficiently utilized for uplink data transmissions and cannot be reduced via reducing $\tau_{0}$ as in \textbf{P4}. This is attributed to the fact that, under \textbf{P3}, the user closer to the BS is also an energy harvesting node which harvests its energy during the same $\tau_{0}$ fraction of time. This, in turn, brings an interesting insight, and may be somewhat surprising at the first glance, that more realistic WPCNs with heterogeneous nodes outperform (in terms of the average maximum sum throughput) WPCNs with energy harvesting nodes only, assuming both are subject to the same overall system constraints.

%In Fig. \ref{fig:2}, Jain's fairness index \cite{13} is plotted for the three systems under consideration against the pathloss exponent considering the same scenario in Fig. \ref{fig:2}. It is observed that the fairness index of both proposed schemes, namely OPIC and OPAC, is less than that of the baseline WPCN with energy harvesting nodes only. This, in turn, highlights one instance of the fundamental throughput-fairness trade-off, where the superior sum throughput performance in \textbf{P1} and \textbf{P3} compared to baseline WPCN came at the expense of a modest degradation in the fairness.
%

 \begin{figure}
\centering
\includegraphics[width=9 cm, height= 7cm]{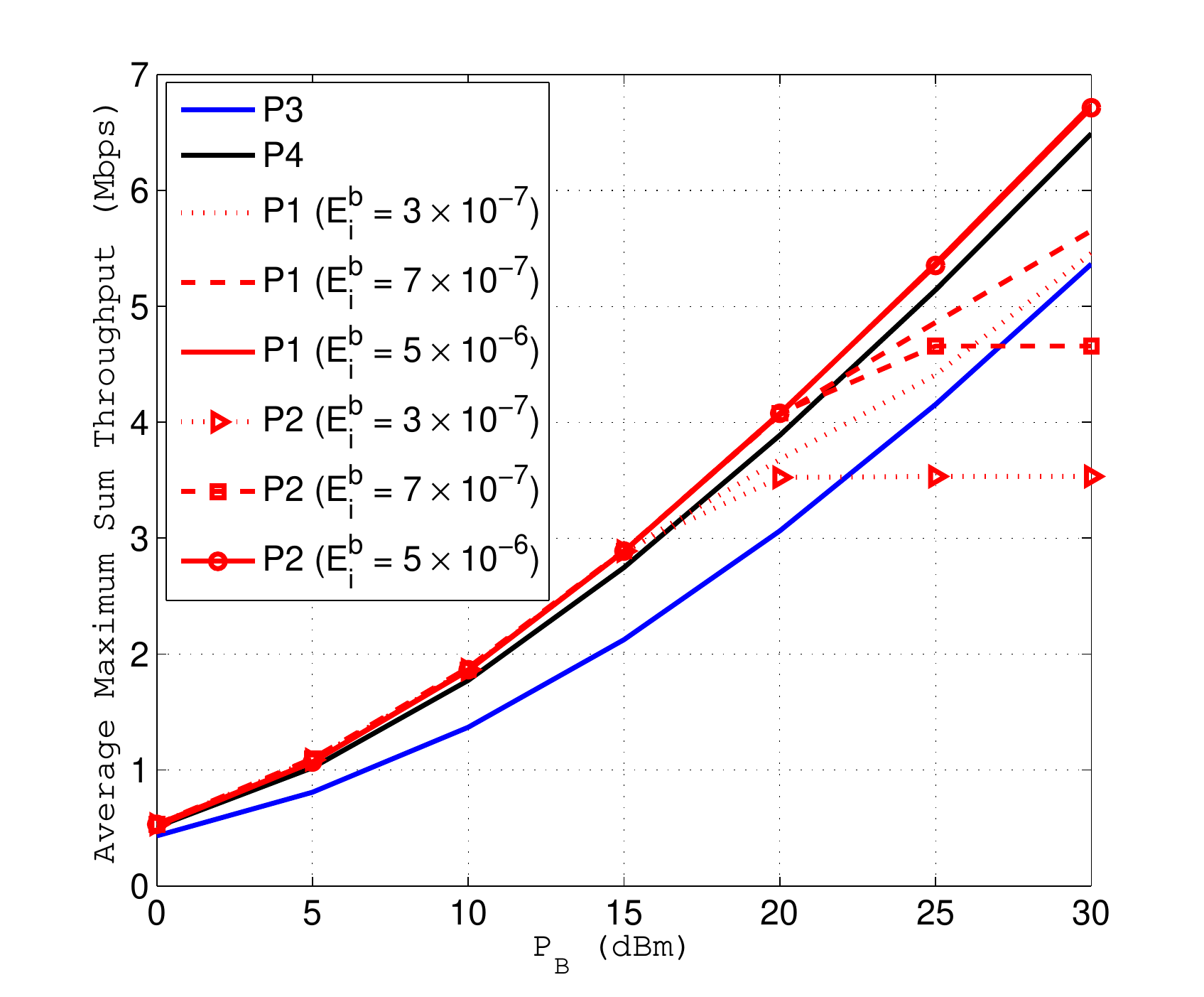}
    \caption{Average maximum sum throughput for all systems with two nodes vs. the BS power, $P_{B}$.}
     \label{fig:6}
\end{figure}

In Fig. \ref{fig:6}, the average maximum sum throughput is plotted for the four systems under consideration against the BS power, $P_{B}$, considering the same scenario in Fig. \ref{fig:5} and using $\beta = 2$. We note that the average maximum sum throughput of the four systems monotonically increases as $P_{B}$ increases. This is intuitive since the average amount of harvested energy by both users in WPCNs with only RF energy harvesting nodes (\textbf{P3}) increases with $P_{B}$. Therefore, $E_{max}$ (the average amount of harvested energy in \textbf{P3}) in \textbf{P1}, \textbf{P2} and \textbf{P4} increases with $P_{B}$. This naturally results in a higher average maximum sum throughput. It is observed that the average maximum sum throughput attained by \textbf{P1} and \textbf{P2}, for the used values of $E_{i}^b$, is the same when $P_{B} \leq 15$ dBm. This is attributed to the fact that if $P_{B} \leq 15$ dBm, the average amount of harvested energy in \textbf{P3} ($E_{max}$ in \textbf{P1}, \textbf{P2} and \textbf{P4}) is very low that \textbf{P1} achieves the average maximum sum throughput via allocating all the entire slot duration for uplink data transmissions (no need for harvesting energy). As $P_{B}$ increases, i.e., $P_{B} > 15$ dBm, the average amount of harvested energy in \textbf{P3} becomes larger, and, hence the RF energy harvesting capability would have a great impact on the performance attained by \textbf{P1}. Therefore, we note that \textbf{P1} outperform \textbf{P2} in terms of the achievable average maximum sum throughput when $E_{i}^{b} = 3 \times 10^{-7} \; \text{and}\; 7 \times 10^{-7}$ Joules. In addition, the average maximum sum throughput of \textbf{P2} saturates.
 \begin{figure}[t!]
\centering
\includegraphics[width=9 cm, height= 7cm]{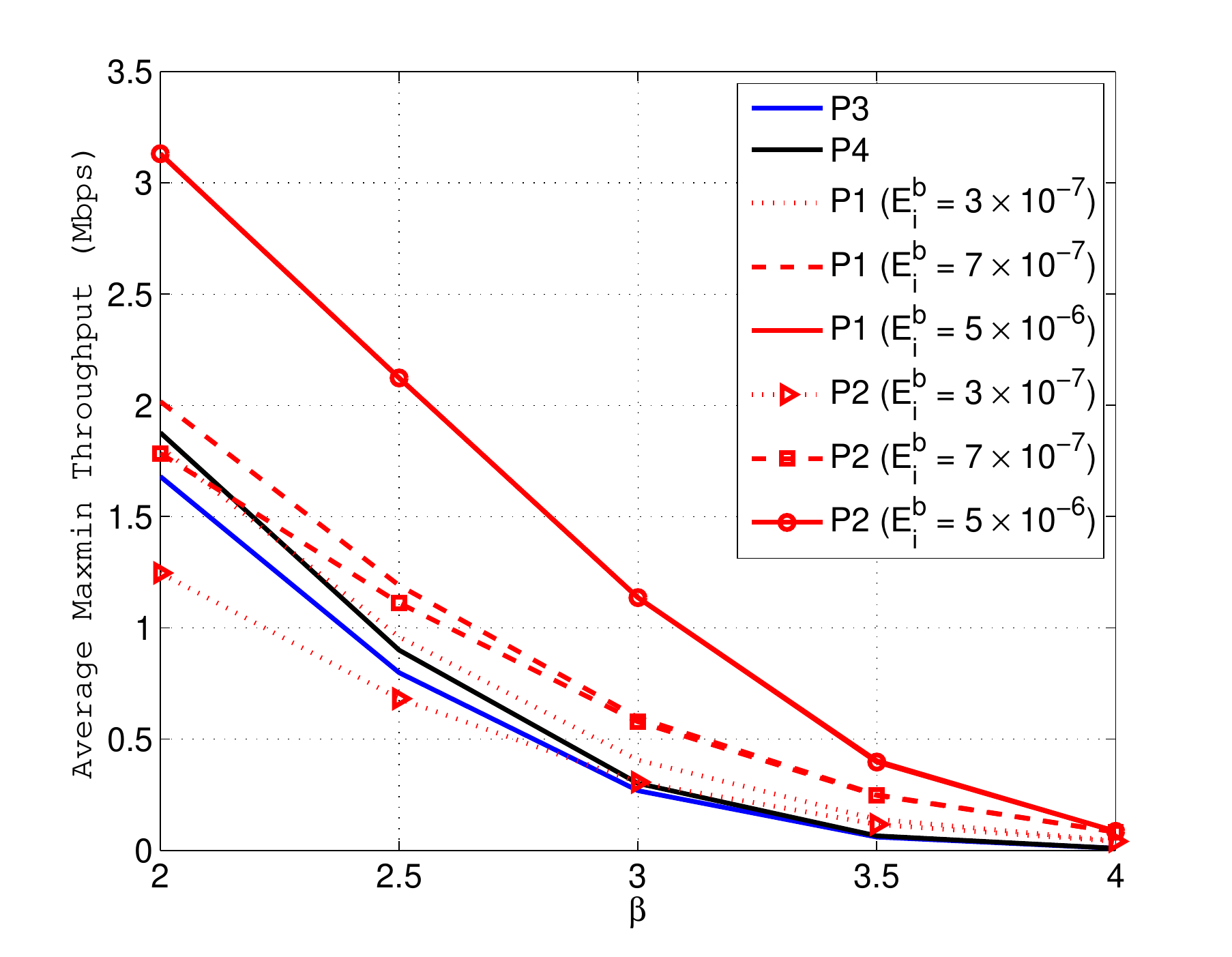}
    \caption{Average maxmin throughput for all systems with two nodes vs. the Pathloss Exponent, $\beta$.}
     \label{fig:7}
\end{figure}
Motivated by the inherent unfairness witnessed for the sum throughput maximization formulation for the four studied systems, Fig. \ref{fig:7} shows the average maxmin throughput comparison with the same set of parameters as in Fig. \ref{fig:5}. First, it is noticed that the average maxmin throughput attained by the generalized formulation ($\textbf{P1}^{\text{Maxmin}}$), for the used values of $E^b_{i}$, along with the conventional TDMA-based wireless network, for $E_{i}^{b} = 3 \times 10^{-7} \; \text{and}\; 7 \times 10^{-7}$, and WPCNs with two types of nodes, all outperform the performance of the WPCN with energy harvesting nodes only. It is also observed that twice the average maxmin throughput of each system (which is the average sum throughput given that we have only two users based on Theorem~\ref{th:6}), at each pathloss exponent value, is less than the average maximum sum throughput for the same system (Fig. \ref{fig:5}). This, in turn, demonstrates the fundamental trade off between achieving maximum sum throughput and achieving fair throughout allocations among different users.

\begin{figure}
\centering
\includegraphics[width=9 cm, height= 7cm]{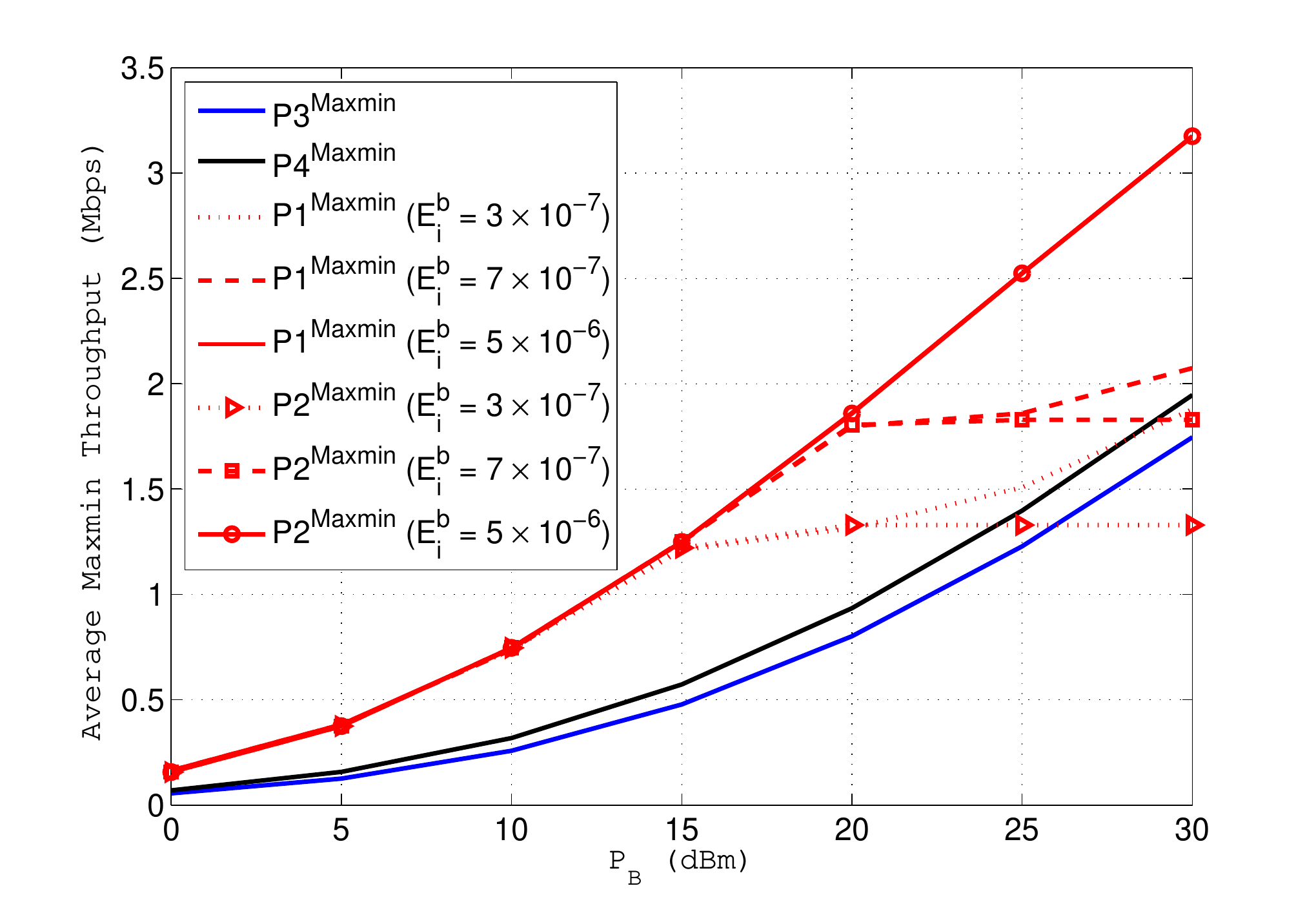}
    \caption{Average maxmin throughput for all systems with two nodes vs. the BS power, $P_{B}$.}
     \label{fig:8}
\end{figure}

In Fig. \ref{fig:8}, the average maxmin throughput is plotted for the four systems against $P_{B}$ considering the same scenario in Fig. \ref{fig:6}. It is observed that the average maxmin throughput of the four systems monotonically increases with $P_{B}$. For small values of $P_{B}$, i.e., $P_{B} \leq 10$ dBm, the average maxmin throughput attained by $\textbf{P1}^{\text{Maxmin}}$ and $\textbf{P2}^{\text{Maxmin}}$, for different values of $E^b_{i}$, achieve the highest average maxmin throughput. In addition, the range of $P_{B}$ values, over which the performance of $\textbf{P2}^{\text{Maxmin}}$ closely follows the performance of $\textbf{P1}^{\text{Maxmin}}$, expands as $E^b_{i}$ increases.

\begin{figure}
\centering
\includegraphics[width=9 cm, height= 7cm]{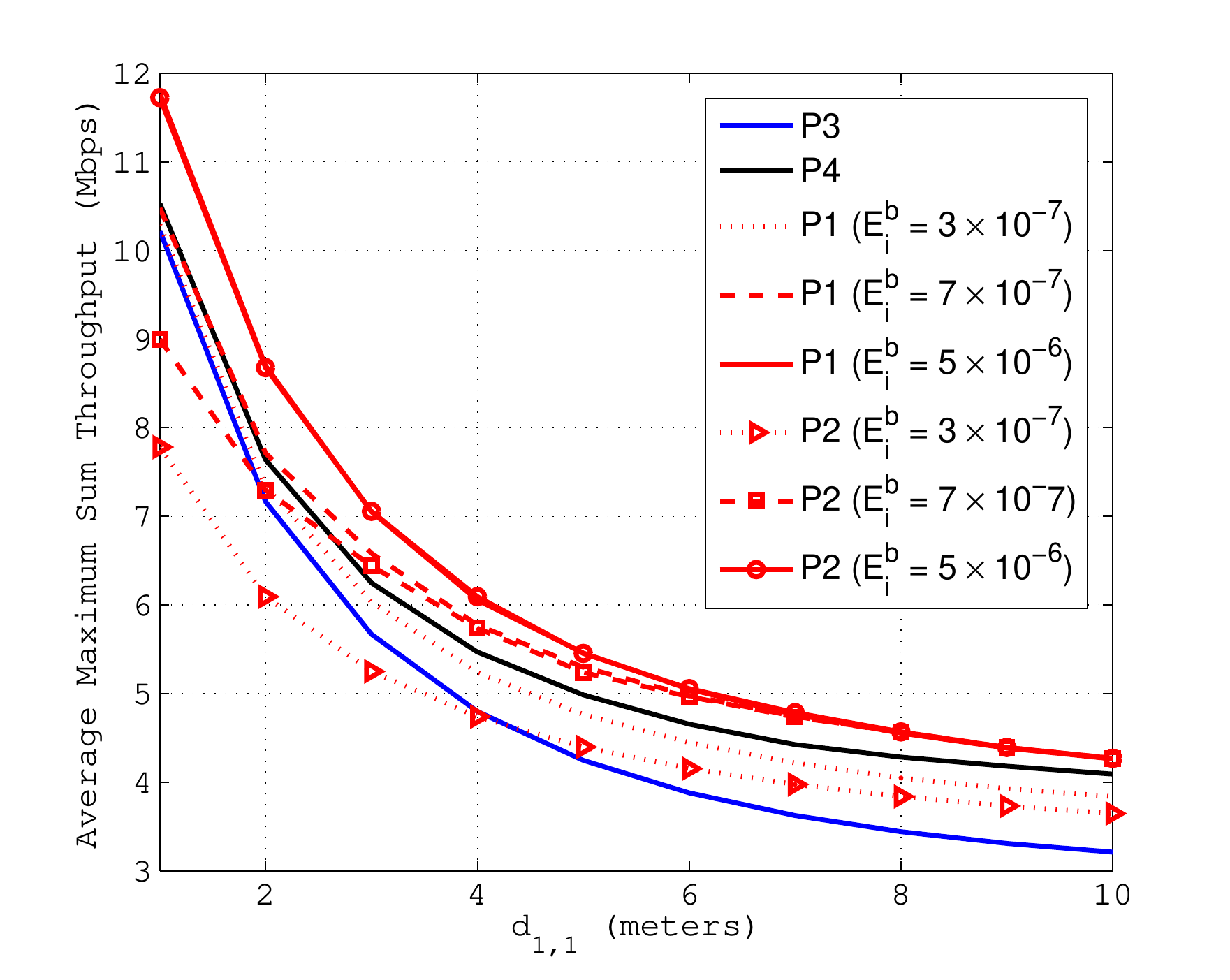}
    \caption{Average maximum sum throughput for all systems with two nodes vs. $U_{1}$'s distance, $d_{1,1}$.}
     \label{fig:10}
\end{figure}

In Fig. \ref{fig:10}, our objective is to emphasize the impact of users' distances, $d_{1,1}$ and $d_{2,1}$, from the BS on the network performance. Towards this objective, the average maximum sum throughput of the four systems under consideration is plotted against $d_{1,1}$. Furthermore, we fix $d_{2,1} = 5$ meters, $\beta = 2$ and $P_{B} = 20$ dBm. We note that the average maximum sum throughput of the four studied systems monotonically decreases as $d_{1,1}$ increases. This is due to the fact that as $d_{1,1}$ increases, $U_{1,1}$ experiences a worse channel in both the uplink and the downlink, and, thus harvests less energy from the BS and requires more energy for uplink data transmissions. Furthermore, similar to Fig. \ref{fig:5} and Fig. \ref{fig:6}, we observe that the average maximum sum throughput attained by \textbf{P2} and \textbf{P3} constitute lower bounds on the performance attained by the generalized setting in \textbf{P1}.

\begin{figure}
\centering
\includegraphics[width=9 cm, height= 7cm]{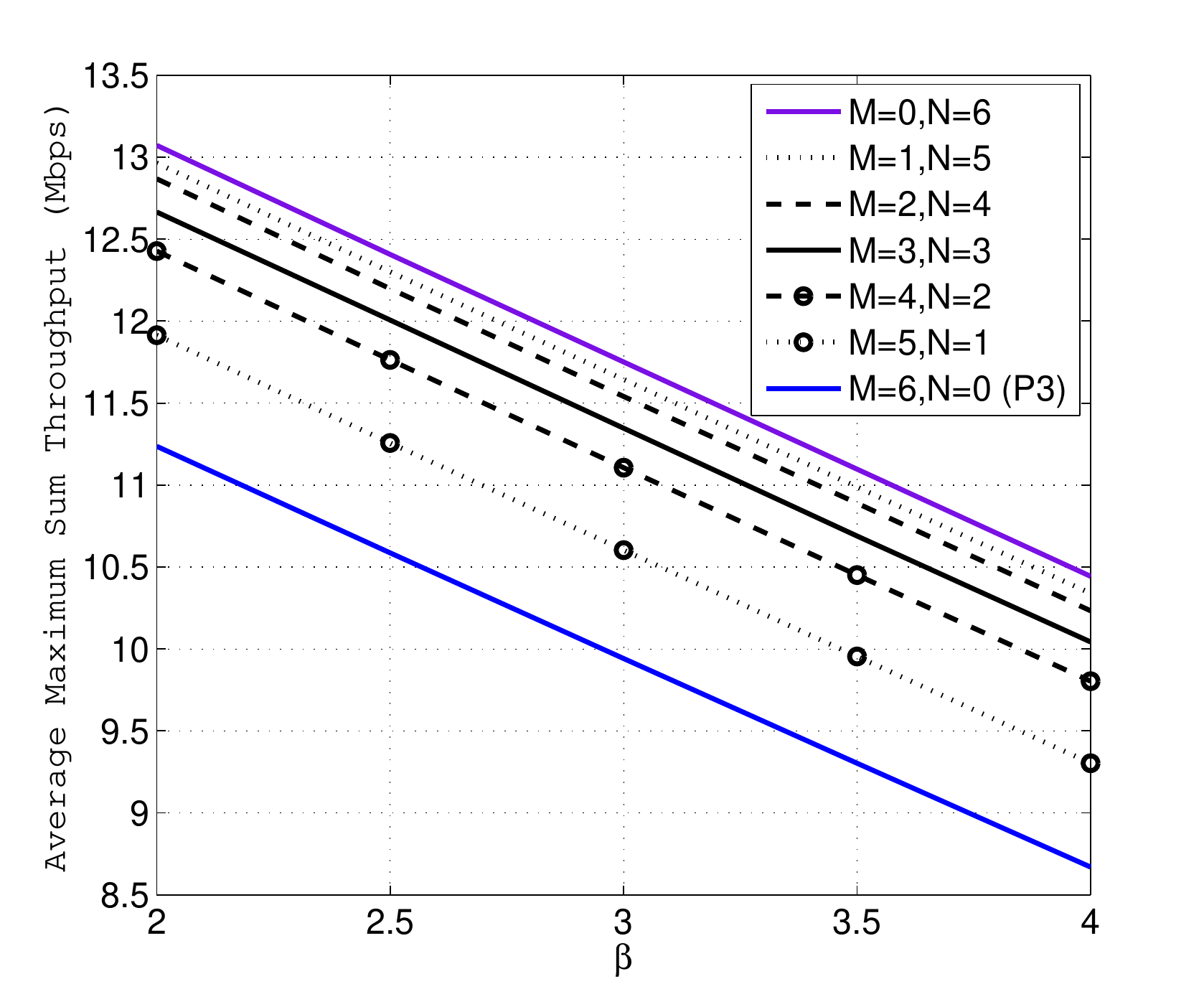}
    \caption{ Average maximum sum throughput for $\textbf{P4}$ vs. the Pathloss Exponent, $\beta$, for different mixes of node types.}
     \label{fig:9}
\end{figure}

Fig. \ref{fig:9} shows the impact of replacing a number of Type I nodes with Type II nodes on WPCNs with heterogeneous nodes performance (\textbf{P4}), via comparing the average maximum sum throughput of $\textbf{P4}$ for different combinations of $M$ and $N$. Towards this objective, we consider a network with six users with same distance $d = \frac{10}{6}$ meters. Note that the insight revealed in Fig. \ref{fig:9} remains valid for all different scenarios with randomized sets of users' distances as demonstrated in Fig. \ref{fig:5} and Fig. \ref{fig:10}. Thus, we focus on the scenario of all users with the same distance to emphasize that effect on the network performance. In addition, we use $P_{B} = 20$ dBm. It is observed that as the number of Type II nodes ($N$) increases, the average maximum sum throughput increases since increasing $N$ reduces the allocated time for energy harvesting $(\tau_{0})$ and, hence, the average maximum sum throughput increases via assigning that reduction in $(\tau_{0})$ for uplink data transmission through Type II nodes. Therefore, it is clear that the highest and lowest average maximum sum throughput are obtained by the extreme cases of $N=6$, $M=0$ and $N=0$, $M=6$ (\textbf{P3}), respectively, as shown in the figure.

% \begin{figure}
% \centering
%\includegraphics[width=9 cm, height= 7cm]{longtermaverageenergy.eps}
%    \caption{Average maximum sum throughput for the 3 systems.}
%     \label{fig:4}
%\end{figure}

%\begin{figure}
%\centering
%\includegraphics[width=9 cm, height= 7cm]{jainfairnessindex.eps}
%    \caption{Jain's fairness index for the 3 systems.}
%     \label{fig:4}
%\end{figure}

%\begin{figure}
%\centering
%\includegraphics[width=9 cm, height= 7cm]{Maxminthroughputaverageenergy.eps}
%    \caption{Average maxmin throughput for the 3 systems.}
%     \label{fig:5}
%\end{figure}

%\begin{figure}
%\includegraphics[width=10 cm, height= 7cm]{effectofEprime.eps}
%    \caption{Throughput vs. $\bar{E}$.}
%     \label{fig:5}
%\end{figure}

%\begin{figure}
%\centering
%\includegraphics[width=9 cm, height= 7cm]{throughputregion.eps}
%    \caption{ Throughput region.}
%     \label{fig:7}
%\end{figure}

\section{Conclusion}
This paper introduces a new, more realistic wireless network setting, coined generalized wireless powered communication networks. Under this setting, each node has two energy sources; a constant energy supply and an RF energy harvesting circuitry. We formulate two optimization problems to investigate the maximum sum throughput and the maxmin throughput. Moreover, we show that different known wireless networks fall as special cases of the proposed system model, namely, conventional TDMA-based wireless networks, WPCNs with only RF energy harvesting nodes and WPCNs with heterogeneous nodes. Our numerical results highlight the great impact of the RF energy harvesting capability on the generalized problem performance, compared to conventional TDMA-based wireless networks. Furthermore, they reveal that the performance of the generalized problem approaches the performance of conventional TDMA-based wireless networks as the amount of allowable consumed energy from constant supply per slot increases. They also demonstrate the fundamental trade off between achieving maximum sum throughput and achieving fairness among different users. In addition, the results reveal the superiority of WPCNs with heterogeneous nodes compared to traditional WPCNs with RF energy harvesting nodes only. As part of the future work, we would like to extend the current framework to multiple BSs.
\label{sec:con}
\section*{Appendix A}
Thanks to the fact that the perspective function of a concave function is also a concave function \cite{35}. $\tau_{i} \log_{2} \left(1 + \alpha_{i} \dfrac{E_{i}} {\tau_{i}}\right)$ is the perspective function of the concave function $\log_{2} \left(1 + \alpha_{i} E_{i}\right)$ which preserves the concavity of $R_{i}$ with respect to $(E_{i},\tau_{i})$. Since the non-negative weighted sum of concave functions is also concave \cite{35}, then the objective function of $\textbf{P1}$, which is the non-negative weighted summation of concave functions, i.e., $R_{i}$ for $i=1, \cdots,K$, is a concave function in $(\mathbf{E},\pmb{\tau})$. In addition, all constraints of $\textbf{P1}$ are affine in $(\mathbf{E},\pmb{\tau})$. This establishes the proof.

\section*{Appendix B}
For a given $\mathbf{E}$ that satisfies (\ref{eq6a}) and $0 \leq E_{i} < E^b_{i} + \eta_{i} P_{B} h_{i}, i=1, \cdots,K$, \textbf{P1} reduces as follows.
\begin{align}
\nonumber & \textbf{P1}^{\prime}:   \nonumber&& \underset{\pmb{\tau}}{\text{max}}\;\;  \sum_{i=1}^{K}{\tau_{i} \log_{2} \left(1 + \alpha_{i} \dfrac{E_{i}}{\tau_{i}}\right)} \\
\label{eq70a}&\text{s.t.} &&\sum_{i=1}^{K}{\tau_{i}} \leq 1 - \tau_{0},  \\
\label{eq70b} &&& \pmb{\tau} \succeq \mathbf{0},\\
\label{eq70c} &&& \tau_{0} \geq \dfrac{E_{i} - E^b_{i}}{\eta_{i} P_{B} h_{i}},\hspace{0.5 cm} i=1, \cdots,K.
\end{align}
It can be easily shown that $R_{i} = \tau_{i} \log_{2} \left(1 + \alpha_{i} \dfrac{E_{i}}{\tau_{i}}\right)$ is a monotonically increasing function in $(E_{i}, \tau_{i})$ [25, Lemma 3.2], $i=1,\cdots,K$. Therefore, the constraint in (\ref{eq70a}) should hold with equality at the optimality (otherwise, the objective function can be further increased by increasing some $\tau_{i}$'s). Hence, from (\ref{eq70c}), the optimal harvesting time duration is given by
\begin{equation}\label{eq71}
\tau_{0}^{*} =  \text{min} \left[ \left(\underset{i}{\text{max}}\lbrace\dfrac{E_{i} - E^b_{i}}{\eta_{i} P_{B} h_{i}}\rbrace\right)^{+},\; 1 \right].
\end{equation}
Hence, $\textbf{P1}^{\prime}$ reduces to
\begin{align}
\nonumber & \textbf{P1}^{\prime \prime}: \hspace{0.5 cm}  \nonumber&& \underset{\pmb{\tau^{\prime}}}{\text{max}}\;\;  \sum_{i=1}^{K}{\tau_{i} \log_{2} \left(1 + \alpha_{i} \dfrac{E_{i}}{\tau_{i}}\right)} \\
\label{eq72a}&\text{s.t.} &&\sum_{i=1}^{K}{\tau_{i}} = 1 - \tau_{0}^{*},  \\
\label{eq72b} &&& \pmb{\tau^{\prime}} \succeq \mathbf{0}. 
\end{align}
Recall that $\pmb{\tau^{\prime}}=[\tau_{1}, \cdots,\tau_{K}]$. Based on Theorem \ref{th:1}, $\textbf{P1}^{\prime \prime}$ is a convex optimization problem and its Lagrangian is given by 
 \begin{dmath} \label{eq73}
 \mathcal{L}\left(\pmb{\tau^{\prime}},\mu\right) = R_{sum}\left(\pmb{\tau^{\prime}}\right) + \mu \left(\sum_{i=1}^{K}{\tau_{i}}- \left(1 - \tau_{0}^{*}\right)\right) ,
 \end{dmath}
 where $R_{sum}\left(\pmb{\tau^{\prime}}\right) = \sum_{i=1}^{K}{\tau_{i} \log_{2} \left(1 + \alpha_{i} \dfrac{E_{i}}{\tau_{i}}\right)} $ and $\mu$ is the Lagrangian dual variable associated with the total slot duration constraint (\ref{eq72a}). It can be easily shown that there exists a $\pmb{\tau^{\prime}}$ that strictly satisfies all constraints of $\textbf{P1} ^{\prime \prime}$. Hence, according to Slater's condition \cite{35}, strong duality holds for this problem; therefore, the KKT conditions are necessary and sufficient for the global optimality of $\textbf{P1} ^{\prime \prime}$, which are given by
 \begin{align} \label{eq74}
 \dfrac{\partial}{\partial \tau_{i}^{*}} \mathcal{L}\left(\pmb{\tau^{\prime *}},\mu^{*}\right) = \log_{2} \left(1 + \alpha_{i} \dfrac{E_{i}}{\tau_{i}^{*}}\right) - \dfrac{\alpha_{i} \dfrac{E_{i}}{ \tau_{i}^{*}}}{\ln(2)\left(1 + \alpha_{i} \dfrac{E_{i}}{\tau_{i}^{*}}\right)} = - \mu^{*},
 \end{align}
$i=1, \cdots,K,$
 \begin{equation} \label{eq75}
 \sum_{i=1}^{K}{\tau_{i}^{*}} = 1 - \tau_{0}^{*},
 \end{equation}
 where $\pmb{\tau^{\prime *}}$ and $\mu^{\ast}$ denote, respectively, the optimal primal and dual solutions of $\textbf{P1}^{\prime \prime}$. Therefore, from (\ref{eq74}) and (\ref{eq75}), we have
 \begin{equation}  \label{eq76}
\alpha_{1} \dfrac{E_{1}}{\tau_{1}^{*}} = \alpha_{2} \dfrac{E_{2}}{\tau_{2}^{*}} \cdots  \alpha_{K} \dfrac{E_{K}}{\tau_{K}^{*}} = \dfrac{\sum_{j=1}^{K}{\alpha_{i} E_{i}}}{1 - \tau_{0}^{*}}.
\end{equation}
Thus from (\ref{eq76}), the optimal time allocations are given by
 \begin{equation} \label{eq78}
\tau_{i}^{*}=\dfrac{\alpha_{i} E_{i} \left(1 - \tau_{0}^{*}\right)}{\sum_{j=1}^{K}{\alpha_{j} E_{j}}},\; i= 1, \cdots,K.
\end{equation}
This establishes the proof.

\section*{Appendix C}
For a given $\pmb{\tau}$ that satisfies (\ref{eq6b}) - (\ref{eq6d}), \textbf{P1} reduces as follows.
\begin{align}
\nonumber & \textbf{P1}^{\dagger}: \hspace{0.5 cm}  \nonumber&& \underset{\mathbf{E}}{\text{max}}\; \; \sum_{i=1}^{K}{\tau_{i} \log_{2} \left(1 + \alpha_{i} \dfrac{E_{i}}{\tau_{i}}\right)} \\
\label{eq79a}&\text{s.t.} &&\sum_{i=1}^{K}{E_{i}} \leq E_{max},  \\
\label{eq79b}&&& 0 \leq E_{i} \leq E^b_{i} + \eta_{i} P_{B} h_{i} \tau_{0},\hspace{0.5 cm} i=1, \cdots,K.
\end{align}
Recall that $R_{i} = \tau_{i} \log_{2} \left(1 + \alpha_{i} \dfrac{E_{i}}{\tau_{i}}\right)$ is a monotonically increasing function in $(E_{i}, \tau_{i})$, $i=1,\cdots,K$. Therefore, when $E_{max} \geq \sum_{j = 1}^{K}{\left(E^b_{j} + \eta_{j} P_{B} h_{j} \tau_{0}\right)}$, $\textbf{P1}^{\dagger}$ has a trivial solution that $E_{i}^{*} = E^b_{i} + \eta_{i} P_{B} h_{i} \tau_{0}$, $i=1, \cdots,K$. On the other hand, when $E_{max} < \sum_{j = 1}^{K}{\left(E^b_{j} + \eta_{j} P_{B} h_{j} \tau_{0}\right)}$, the optimal solution of $\textbf{P1}^{\dagger}$ can be characterized as follows. First, the constraint in (\ref{eq79a}) should hold with equality at the optimality (otherwise, the objective function can be further increased by increasing some $E_{i}$'s). Based on Theorem \ref{th:1}, $\textbf{P1}^{\dagger}$ is a convex optimization problem and its Lagrangian is given by 
 \begin{equation} \label{eq80}
 \mathcal{L}\left(\mathbf{E},\lambda\right) = R_{sum}\left(\mathbf{E}\right) + \lambda \left(\sum_{i=1}^{K}{E_{i}} - E_{max}\right) ,
 \end{equation}
 where $\lambda$ is the Lagrangian dual variable associated with the total allowable consumed energy per slot constraint (\ref{eq79a}). The strong duality holds for $\textbf{P1}^{\dagger}$; therefore, the KKT conditions are necessary and sufficient for the global optimality of $\textbf{P1}^{\dagger}$, which are given by
 \begin{equation} \label{eq81}
 \dfrac{\partial}{\partial E_{i}^{*}} \mathcal{L}\left(\mathbf{E}^{*},\lambda^{\ast}\right) = \dfrac{\alpha_{i}}{\ln(2)\left(1 + \dfrac{\alpha_{i} E_{i}^{*}}{\tau_{i}}\right)} + \lambda^{*}= 0,\; i=1, \cdots,K,
 \end{equation}
 \begin{equation} \label{eq82}
 \sum_{i=1}^{K}{E_{i}^{*}} = E_{max},
 \end{equation}
 where $\mathbf{E}^{*}$ and $\lambda^{\ast}$ denote, respectively, the optimal primal and dual solutions of $\textbf{P1}^{\dagger}$. Therefore, from (\ref{eq81}), we have
 \begin{equation} \label{eq83}
E_{i}^{*} = - \dfrac{\tau_{i}}{\alpha_{i}}\left(\dfrac{\alpha_{i}}{\lambda^{*} \ln(2)} + 1\right),\; i=1, \cdots,K.
 \end{equation}
Taking into account the constraints in (\ref{eq79b}), the optimal energy allocations are given by
 \begin{equation}\label{eq85}
E_{i}^{*} = \text{min}\left[\left(- \dfrac{\tau_{i}}{\alpha_{i}}\left(\dfrac{\alpha_{i}}{\lambda^{*} \ln(2)} + 1\right)\right)^{+},\;E^{b}_{i} + \eta_{i} P_{B} h_{i} \tau_{0} \right], 
 \end{equation}
where $i=1,\cdots,K$ and $\lambda^{*}$ satisfies the equality constraint $\sum_{i=1}^{K}{E_{i}^{*}} = E_{max}$.  This establishes the proof.

\section*{Appendix D}
 \textbf{P4} is a convex optimization problem and its Lagrangian is given by
 \begin{dmath}  \label{eq34}
 \mathcal{L}\left(\pmb{\tau^{\prime \prime}},\bar{E},\lambda,\mu\right) = R_{sum}\left(\pmb{\tau^{\prime \prime}},\bar{E}\right) - \mu \left(\tau_{0}+\sum_{i=1}^{M}{\tau_{1,i}}+\sum_{j=1}^{N}{\tau_{2,j}} - 1\right) - \lambda \left(a\tau_{0} + N \bar{E} - E_{max}\right),
 \end{dmath}
where $\mu$ and $\lambda$ are the Lagrangian dual variables associated with the slot duration and the total allowable consumed energy per slot constraints, respectively, and $R_{sum}\left(\pmb{\tau^{\prime \prime}},\bar{E}\right) = \sum_{i=1}^{M}{R_{1,i}\left(\tau_{0},\tau_{1,i}\right)} + \sum_{j=1}^{N}{R_{2,j}\left(\bar{E},\tau_{2,j}\right)}$. Hence, the dual function can be expressed as
\begin{equation}  \label{eq35}
G\left(\lambda,\mu\right) = \underset{\pmb{\tau^{\prime \prime}},\bar{E}\in \mathcal{S}}{\max } \; \mathcal{L}\left(\pmb{\tau^{\prime \prime}},\bar{E},\lambda,\mu\right),
\end{equation}
where $\mathcal{S}$ is the feasible set specified by $\pmb{\tau^{\prime \prime}} \succeq \mathbf{0}$ and $\bar{E} \geq 0$. It can be easily shown that there exists a $(\pmb{\tau^{\prime \prime}},\bar{E})$ that strictly satisfies all constraints of \textbf{P4}. Hence, according to Slater's condition \cite{35}, strong duality holds for this problem; therefore, the KKT conditions are necessary and sufficient for the global optimality of \textbf{P4}, which are given by
\begin{dmath}  \label{eq36}
\tau_{0}^{\ast}+\sum_{i=1}^{M}{\tau_{1,i}^{\ast}}+\sum_{j=1}^{N}{\tau_{2,j}^{\ast}} \leq 1,
\end{dmath} 
\begin{equation} \label{eq37}
a\tau_{0}^{\ast} + N \bar{E}^{\ast} \leq E_{max},
\end{equation}
\begin{equation}  \label{eq38}
\mu^{\ast}  \left(\tau_{0}^{\ast}+\sum_{i=1}^{M}{\tau_{1,i}^{\ast}}+\sum_{j=1}^{N}{\tau_{2,j}^{\ast}} - 1\right) = 0,
\end{equation}
\begin{equation} \label{eq39}
\lambda^{\ast} \left(a\tau_{0}^{\ast} + N \bar{E}^{\ast} - E_{max}\right) = 0,
\end{equation}
\begin{equation}  \label{eq40}
\dfrac{\partial}{\partial \tau_{0}}R_{sum}\left(\pmb{\tau^{\prime \prime \ast}},\bar{E}^{\ast}\right) - \left(a \lambda^{\ast} + \mu^{\ast}\right) = 0,
\end{equation}
\begin{equation}  \label{eq41}
\dfrac{\partial}{\partial \tau_{1,i}}R_{sum}\left(\pmb{\tau^{\prime \prime \ast}},\bar{E}^{\ast}\right) - \mu^{\ast} = 0, \; i=1, \cdots,M,
\end{equation}
\begin{equation}  \label{eq42}
\dfrac{\partial}{\partial \tau_{2,j}}R_{sum}\left(\pmb{\tau^{\prime \prime \ast}},\bar{E}^{\ast}\right) - \mu^{\ast} = 0,\; j=1, \cdots,N,
\end{equation}
\begin{equation} \label{eq43}
\dfrac{\partial}{\partial \bar{E}^{\ast}}R_{sum}\left(\pmb{\tau^{\prime \prime \ast}},\bar{E}^{\ast}\right) - N \lambda^{\ast} = 0,
\end{equation}
where $\left(\pmb{\tau^{\prime \prime \ast}},\bar{E}^{\ast}\right)$ and $\left(\lambda^{\ast},\mu^{\ast}\right)$ denote, respectively, the optimal primal and dual solutions of \textbf{P4}. Since $R_{sum}\left(\pmb{\tau^{\prime \prime}},\bar{E}\right)$ is a monotonic increasing function in $\left(\pmb{\tau^{\prime \prime}},\bar{E}\right)$, therefore $\tau_{0}^{\ast}+\sum_{i=1}^{M}{\tau_{1,i}^{\ast}}+\sum_{j=1}^{N}{\tau_{2,j}^{\ast}} = 1$ and $a\tau_{0}^{\ast} + N \bar{E}^{\ast} = E_{max}$ must hold. From (\ref{eq40}) - (\ref{eq43}), we have
\begin{equation}  \label{eq44}
\sum_{i=1}^{M}{\dfrac{\gamma_{i}}{1 + \gamma_{i} \dfrac{\tau_{0}^{\ast}}{\tau_{1,i}^{\ast}}}} = \left(a \lambda^{\ast} + \mu^{\ast}\right) \ln(2),
\end{equation}
\begin{equation}  \label{eq45}
\ln\left(1 + \gamma_{i} \dfrac{\tau_{0}^{\ast}}{\tau_{1,i}^{\ast}}\right) - \dfrac{\gamma_{i}\dfrac{\tau_{0}^{\ast}}{\tau_{1,i}^{\ast}}}{1+\gamma_{i}\dfrac{\tau_{0}^{\ast}}{\tau_{1,i}^{\ast}}} = \mu^{\ast} \ln(2),\; i=1, \cdots,M,
\end{equation}
\begin{equation}  \label{eq46}
\ln\left(1+\dfrac{\bar{E}^{\ast} \theta_{j}}{\tau_{2,j}^{\ast}}\right) - \dfrac{\dfrac{\bar{E}^{\ast} \theta_{j}}{\tau_{2,j}^{\ast}}}{1 + \dfrac{\bar{E}^{\ast} \theta_{j}}{\tau_{2,j}^{\ast}}} = \mu^{\ast} \ln(2), \; j=1, \cdots, N.
\end{equation}
\begin{equation} \label{eq47}
\sum_{j=1}^{N}{\dfrac{\theta_{j}}{1 + \theta_{j} \dfrac{\bar{E}^{\ast}}{\tau_{2,j}^{\ast}}}} = N\lambda^{\ast}\ln(2),
\end{equation}
Therefore, from (\ref{eq45}) and (\ref{eq46}), we have
\begin{equation}  \label{eq48}
\dfrac{\gamma_{1} \tau_{0}^{\ast}}{\tau_{1,1}^{\ast}} = \dfrac{\gamma_{2} \tau_{0}^{\ast}}{\tau_{1,2}^{\ast}} = \cdots  \dfrac{\gamma_{M} \tau_{0}^{\ast}}{\tau_{1,M}^{\ast}} = \dfrac{\bar{E}^{\ast}\theta_{1}}{\tau_{2,1}^{\ast}} = \dfrac{\bar{E}^{\ast}\theta_{2}}{\tau_{2,2}^{\ast}} = \cdots  \dfrac{\bar{E}^{\ast}\theta_{N}}{\tau_{2,N}^{\ast}} = x_{1}.
\end{equation}
From $\tau_{0}^{\ast}+\sum_{i=1}^{M}{\tau_{1,i}^{\ast}}+\sum_{j=1}^{N}{\tau_{2,j}^{\ast}} = 1$ and (\ref{eq48}), $\tau_{1,i}^{\ast}$ and $\tau_{2,j}^{\ast}$ can be expressed, respectively, by

\begin{equation}  \label{eq49}
\tau_{1,i}^{\ast} = \dfrac{\gamma_{i} \left(N\left(x_{1}^{\ast}- 1\right) - E_{max} A_{2} \right) }{\left(x_{1}^{\ast} - 1\right) \left(N\left(x_{1}^{\ast} - 1 + A_{1}\right) - a A_{2} \right)}, \;  i=1, \cdots, M,
\end{equation}
\begin{equation}  \label{eq50}
\tau_{2,j}^{\ast} = \dfrac{\theta_{j} \left(E_{max}\left(x_{1}^{\ast} - 1 + A_{1} \right) - a \left(x_{1}^{\ast} - 1\right)\right) }{\left(x_{1}^{\ast} - 1\right)\left(N\left(x_{1}^{\ast} - 1 + A_{1}\right) - a A_{2} \right)}, \;  j=1, \cdots, N,
\end{equation}
where $A_{1} = \sum_{i=1}^{M}{\gamma_{i}}$ and $A_{2} = \sum_{j=1}^{N}{\theta_{j}}$. From (\ref{eq44}) and (\ref{eq47}), it follows that
\begin{equation}\label{eq51}
\lambda^{\ast} = \dfrac{A_{2}}{ N x_{1} \ln(2)},
\end{equation}
\begin{equation}\label{eq52}
\mu^{\ast} = \dfrac{A_{1} - \dfrac{a}{N}A_{2}}{x_{1} \ln(2)}.
\end{equation}
By substituting with $\mu^{\ast}$ into (\ref{eq45}), we have
\begin{equation}  \label{eq53}
x_{1}\ln(x_{1}) - x_{1} + 1 = A_{1} - \dfrac{a}{N} A_{2}.
\end{equation}
 
 From (\ref{eq49}) and (\ref{eq50}), it is clear that $x_{1} > 1$ if $A_{1} > 0$,  $A_{2} > 0$ and $0< \tau_{0}^{\ast} < 1$. According to \cite[Lemma 3.2]{21}, there exists a unique solution $x_{1}^{\ast} > 1$ for (\ref{eq53}) if $A_{1} \geq \dfrac{a}{N} A_{2}$, otherwise the total slot time and the total allowable consumed energy per slot will be assigned to the Type II nodes for uplink information transmissions. Thus from (\ref{eq48})-(\ref{eq53}), the optimal time and energy allocations are given by 
\begin{equation}  \label{eq54}
\tau_{0}^{\ast}  = \dfrac{N\left(x_{1}^{\ast} - 1 \right)- E_{max} A_{2}  }{N\left(x_{1}^{\ast} - 1 + A_{1}\right) - a A_{2}},
\end{equation}
\begin{equation}  \label{eq55}
\tau_{1,i}^{\ast} = \dfrac{\gamma_{i} \left(x_{1}^{\ast} - E_{max} A_{2} - 1\right) }{\left(x_{1}^{\ast} - 1\right) \left(x_{1}^{\ast} + A_{1} - a A_{2} -1\right)}, \; i=1, \cdots,M,
\end{equation}
\begin{equation}  \label{eq56}
\tau_{2,j}^{\ast} = \dfrac{\theta_{j} \left(E_{max}\left(x_{1}^{\ast} + A_{1} - 1\right) - a \left(x_{1}^{\ast} - 1\right)\right) }{K\left(x_{1}^{\ast} - 1\right)\left(x_{1}^{\ast} + A_{1} - a A_{2} -1\right)}, \; j=1, \cdots,N.
\end{equation}
\begin{equation} \label{eq57}
\bar{E}^{\ast} =  \dfrac{E_{max}\left(x_{1}^{\ast} - 1 + A_{1} \right) - a \left(x_{1}^{\ast} - 1\right) }{N\left(x_{1}^{\ast} - 1 + A_{1}\right) - a A_{2}}.
\end{equation}

From (\ref{eq54}) - (\ref{eq57}) and taking into account that $[\tau_{0}^{*}, \tau_{1,1}^{*}, \cdots,\tau_{1,M}^{*}, \tau_{2,1}^{*}, \cdots,\tau_{2,N}^{*}, \bar{E}^{*}] \succeq \mathbf{0}$ , we must have $\dfrac{a(x_{1}^{*} - 1)}{A_{1} + x_{1}^{*} - 1} \leq E_{max} \leq \dfrac{N}{A_{2}}(x_{1}^{*} - 1)$. If $E_{max} > \dfrac{N}{A_{2}}(x_{1}^{*} - 1)$, then we have $[\tau_{0}^{*}, \tau_{1,1}^{*}, \cdots,\tau_{1,M}^{*}] \prec \mathbf{0}$. Hence, the total slot time and the total allowable consumed energy per slot will be assigned to the Type II nodes for uplink information transmissions. Therefore, from (\ref{eq38}) and (\ref{eq46}), the optimal time and energy allocations are given by (\ref{eq16})-(\ref{eq19}).
%\begin{equation}  \label{eq58}
%\tau_{0}^{\ast}  = 0,
%\end{equation}
%\begin{equation}  \label{eq59}
%\tau_{1,i}^{\ast} = 0, \; i=1, \cdots,M,
%\end{equation}
%\begin{equation}  \label{eq60}
%\tau_{2,j}^{\ast} = \dfrac{\theta_{j}}{A_{2}}, \; j=1, \cdots,N,
%\end{equation}
%\begin{equation} \label{eq61}
%\bar{E}^{\ast} = \dfrac{E_{max}}{N}.
%\end{equation}

On the other hand, if $E_{max} < \dfrac{a(x_{1}^{*} - 1)}{A_{1} + x_{1}^{*} - 1}$, then we have $[\tau_{2,1}^{*}, \cdots,\tau_{2,N}^{*}, \bar{E}^{*}] \prec \mathbf{0}$. Hence, the total slot time and the total allowable consumed energy per slot will be assigned to the Type I nodes for uplink information transmissions. Therefore, from (\ref{eq38}), (\ref{eq44}) and (\ref{eq45}), the optimal time and energy allocations are given by (\ref{eq16})-(\ref{eq19}).
%\begin{equation}  \label{eq62}
%\tau_{0}^{\ast}  =  \text{min} \left[\dfrac{x^{*} - 1}{A_{1} + x^{*} - 1} , \dfrac{E_{max}}{a}\right],
%\end{equation}
%\begin{equation}  \label{eq63}
%\tau_{1,i}^{\ast} = \text{max} \left[\dfrac{\gamma_{i}}{A_{1} + x^{*} - 1} , \dfrac{\gamma_{i}}{A_{1}}\left(1 - \dfrac{E_{max}}{a}\right)\right], \; i=1, \cdots,M,
%\end{equation}
%\begin{equation}  \label{eq64}
%\tau_{2,j}^{\ast} = 0, \; j=1, \cdots,N,
%\end{equation}
%\begin{equation} \label{eq65}
%\bar{E}^{\ast} = 0.
%\end{equation}
This establishes the proof.
%\bibliography{mypaper}
%\bibliographystyle{IEEEtran}

\end{document}